\documentclass[a4paper,11pt]{article}
\pdfoutput=1
\usepackage[normalem]{ulem}

\usepackage{jcappub} 
                     
\usepackage{amsmath}
\usepackage{amssymb}
\usepackage{color}
\usepackage{graphicx}
\usepackage{comment}
\usepackage{multirow}
\usepackage[dvipsnames]{xcolor}
\usepackage[utf8]{inputenc}
\usepackage{caption}
\usepackage{epstopdf}

\usepackage[T1]{fontenc} 

\newcommand{\HII}{\rm H~{\sc ii}}

\newcommand{\TB}{\delta T_{\rm b}}
\newcommand{\MSUN}{{\rm M}_{\odot}}

\newcommand{\XHI}{x_{\rm HI}}

\newcommand{\TS}{T_{\rm S}}
\newcommand{\TK}{T_{\rm K}}
\newcommand{\TCMB}{T_{\gamma}}
\newcommand{\lya}{\rm {Ly{\alpha}}}
\newcommand{\lyb}{\rm {Ly{\beta}}}
\newcommand{\OmegaB}{\Omega_{\rm B}}
\newcommand{\Omegam}{\Omega_{\rm m}}
\newcommand{\mFDM}{m_{22}}
\newcommand{\mfdm}{m_{\textup{FDM}}}
\newcommand{\aosc}{a_{\textup{osc}}}
\newcommand{\tosc}{t_{\textup{osc}}}

\newcommand{\deltaTb}{\delta {T_{\textup{b}}}}
\newcommand{\deltaB}{\delta_{\textup{B}}}
\newcommand{\deltaAlpha}{\delta_{\alpha}}
\newcommand{\betaAlpha}{\beta_{\alpha}}
\newcommand{\deltaX}{\delta_{\textup{x}}}
\newcommand{\betaX}{\beta_{\textup{x}}}
\newcommand{\deltaT}{\delta_{T}}
\newcommand{\betaT}{\beta_{T}}

\newcommand{\xAlpha}{x_{\alpha}}

\newcommand{\Nion}{N_{\rm ion}}
\newcommand{\NionPopII}{N_{\rm ion, PopII}}
\newcommand{\NionPopIII}{N_{\rm ion, PopIII}}
\newcommand{\fPopII}{f_{\rm PopII}}

\newcommand{\kJeans}{k_{\textup{J,FDM}}}
\newcommand{\kJeansEq}{k_{\textup{Jeq,FDM}}}
\newcommand{\MJeans}{M_{\textup{J,FDM}}}
\newcommand{\MJeansEq}{M_{\textup{Jeq,FDM}}}
\newcommand{\PCDM}{P_{\textup{CDM}}}
\newcommand{\PFDM}{P_{\textup{FDM}}}
\newcommand{\TFDM}{T_{\textup{F}}}
\newcommand{\deltacrit}{\delta_{\textup{crit}}}
\newcommand{\deltacritEdS}{\delta_{\textup{crit,CDM}}}
\newcommand{\GF}{G_{\textup{F}}}
\newcommand{\hF}{h_{\textup{F}}}
\newcommand{\deltavir}{\Delta_{\textup{vir}}}

\newcommand*{\everymodeprime}{\ensuremath{\prime}}

\title{Fuzzy Dark Matter at Cosmic Dawn: New 21-cm Constraints}

\author[a]{Olof Nebrin,}
\author[a]{Raghunath Ghara,}
\author[a]{Garrelt Mellema}

\affiliation[a]{Department of Astronomy and Oskar Klein Centre, \\Stockholm University, AlbaNova, SE-106 91 Stockholm, Sweden}
\emailAdd{olne8343@astro.su.se}
\emailAdd{raghunath.ghara.@astro.su.se}
\emailAdd{garrelt.mellema@astro.su.se}

\abstract{
Potential small-scale discrepancies in the picture of galaxy formation painted by the $\Lambda$CDM paradigm have led to considerations of modified dark matter models. One such dark matter model that has recently attracted much attention is fuzzy dark matter (FDM). In FDM models, the dark matter is envisaged to be an ultra-light scalar field with a particle mass $\mfdm \sim 10^{-22} ~\textup{eV}$. This yields astronomically large de Broglie wavelengths which can suppress small-scale structure formation and give rise to the observed kpc-sized density cores in dwarf galaxies.  We investigate the evolution of the 21-cm signal during Cosmic Dawn and the Epoch of Reionization (EoR) in $\Lambda$FDM cosmologies using analytical models. The delay in source formation and the absence of small halos in $\Lambda$FDM significantly postpone the $\lya$ coupling, heating, as well as the reionization of the neutral hydrogen of the intergalactic medium. 
As a result, the absorption feature in the evolution of the global 21-cm signal has a significantly smaller full width at half maximum ($\Delta z \lesssim 3$), than $\Lambda$CDM ($\Delta z \simeq 6$). This alone rules out $\mfdm < 6 \times 10^{-22} ~\textup{eV}$ as a result of the $2\sigma$ lower limit $\Delta z \gtrsim 4$ from EDGES High-Band.
As a result, $\Lambda$FDM is not a viable solution to the potential small-scale problems facing $\Lambda$CDM. Finally, we show that any detection of the 21-cm signal at redshifts $z > 14$ by interferometers such as the SKA can also exclude $\Lambda$FDM models. }

\keywords{reionization, first stars, X-rays, dark matter theory, axions, power spectrum}

\begin{document}
\maketitle
\flushbottom



\section{Introduction}
\label{intro}

What is the nature of dark matter? When and how did the first stars and galaxies form in the Universe? When and how did the Universe reionize? These are some of the most pressing questions for modern-day cosmology. The cosmological backdrop provided by the $\Lambda$CDM paradigm renders it possible to begin to address them. The development and broad-brush success of the inflationary $\Lambda$CDM paradigm has offered a new synopsis of the evolution of the Universe over large swathes of time and many decades in scale. The main aspects of this paradigm include the inflationary predictions of a flat Universe perturbed by a Gaussian and nearly scale-invariant spectrum of primordial density fluctuations \citep[e.g.][]{Brandenberger1984,Guth1985,Tegmark2005}, a recent accelerating expansion of the Universe driven by something resembling a cosmological constant \citep[][]{Riess1998,Perlmutter1999,Rubin2016}, and the hierarchical assembly of large-scale structure expected in a cold dark matter (CDM) Universe \citep[][]{Blumenthal1984}.

If $\Lambda$CDM accurately describes dark matter on sub-megaparsec (Mpc) scales, the root mean square (RMS) amplitude of linear density fluctuations grows as smaller and smaller scales are surveyed until the free-streaming length is reached. The larger amplitudes on small scales lead to a hierarchical picture of structure formation in which low-mass dark matter halos form earlier and in greater abundance than high-mass halos.

The fate of the baryons in these low-mass halos at high redshifts will be determined by how rapidly the gas can cool and condense \citep[][]{WhiteRees1978, Blumenthal1984}. The very first Population III (PopIII) stars in the $\Lambda$CDM model are expected to form in so-called ``mini-halos'' with masses $\sim 10^{6} ~ \MSUN$ at redshifts $z \sim 20 - 30$ in which cooling by molecular hydrogen  (H$_{2}$) is effective \citep[e.g.][]{Tegmark1997, Bromm2013}. Soon thereafter, this avenue for star formation can be shut down as a result of Lyman-Werner radiation produced by these first stars which can dissociate the H$_{2}$ \citep[][]{Haiman1997}. Without H$_{2}$-driven cooling, star formation is limited to the earliest ``atomic-cooling'' halos with virial temperatures in excess of $\sim 10^4 ~ \rm K$, the limit below which $\lya$ cooling becomes ineffective \citep[e.g.][]{Oh2002, 2007ApJ...665..899W, Greif2008, 2011ApJ...731...54P, Kimm2016}. These halos have characteristic masses $\sim 10^8 ~ \MSUN$, typically collapse at redshifts $z \sim 10 - 20$, and are expected to host the first galaxies. The formation of these halos marks the so-called Cosmic Dawn when the first galactic sources began to have an impact on the state of the intergalactic medium (IGM). This eventually culminates in the Epoch of Reionization (EoR) where the IGM evolves into an ionized state in the presence of strong ionizing UV-sources, completing the process around a redshift $z \sim 6$ \citep[][]{Banados2018}. 

The progress and timing of events just outlined is intimately connected to the nature of dark matter. If the power spectrum of density fluctuations is suppressed on comoving mass-scales $M \lesssim 10^6 ~ \MSUN$, this could substantially reduce the number of PopIII stars forming in minihalos. Stronger suppression of the power spectrum on mass scales $M \lesssim {\rm few} \times 10^8 ~ \MSUN$ could further delay the EoR and the heating of the IGM --- effects that could be observationally probed. 

Such probes have become increasingly urgent in view of the reported small-scale problems with galaxy formation in $\Lambda$CDM. These are the missing satellites problem \citep[e.g.][]{MissingSatellites, Li2018, Smercina2018}, the Too-Big-To-Fail Problem \citep[e.g.][]{Boylan-Kolchin2011, Boylan-Kolchin2012, Papastergis2016}, the Core-Cusp problem \citep[e.g.][]{FloresPrimack1994, Zhu2016, Bose2018, Benitez2018} and the Plane Satellites Problem \citep[e.g.][]{SatellitePlane2018}.
For a recent review mainly relevant for the first three, see \cite{SmallScaleTensionReview2017}. An up-to-date review of the Plane Satellites Problem can be found in \cite{PlanesReview}. Whether these challenges to $\Lambda$CDM are mainly indicative of unaccounted baryonic feedback mechanisms, new exotic dark matter physics on small scales, or a mixture of the two, remains debated. On top of the small-scale challenges to $\Lambda$CDM, Weakly Interactive Massive Particles (WIMPs) --- long heralded as one of the most promising CDM candidates --- have continued to elude the most recent and sensitive experiments \citep[e.g.][]{XENON1T2018}.

These concerns with $\Lambda$CDM models motivate a serious look at alternative dark matter models that could mitigate the small-scale challenges faced by $\Lambda$CDM, and at the same time be unconstrained by recent experimental searches. One such dark matter model, fuzzy dark matter (FDM), sometimes called wave dark matter or $\psi$DM \citep[see e.g.][]{Hu2000,Woo2009, Schive2014Nature, Schive2016, Du2017, FirstStarsFDM}, has gained an increasing amount of attention as of late \citep[for a recent review, see][]{Hui2017}. In FDM models, it is postulated that most of the dark matter has its origin in a new scalar field, the quanta of which (the FDM particles) have exceedingly small masses on the order of $\mfdm \sim 10^{-22} ~ \textup{eV}$. The small particle mass entails astronomically large de Broglie wavelengths which can simultaneously suppress small-scale structure formation and yield kpc-sized cores in dark matter halos. This renders possible the solution of the Missing Satellites Problem, the Core-Cusp Problem, and perhaps the related Too-Big-To-Fail Problem \citep[][]{Robles2018}, in terms of $\mfdm$ --- the only free parameter.

The ``all-in-one'' solution to the small-scale problems facing $\Lambda$CDM provided by $\Lambda$FDM gives it an advantage over warm dark matter (WDM) models. WDM is, unlike FDM, thermally produced in the early Universe, with non-negligible free-streaming that can suppresses small-scale structure formation \citep[e.g.][]{Viel2005}. However, WDM models suffer from a \textit{Catch 22} problem when it comes to producing cores in halo density profiles: A kpc-sized core in a dwarf galaxy translates into a suppression of the power spectrum that would prohibit the formation of the dwarf galaxy in the first place \citep[][]{Catch22}. Thus, WDM cannot simultaneously solve the Missing Satellites Problem and the Cusp-Core Problem. This trap is avoided in FDM because for a given suppression of the power spectrum, FDM generates significantly more extended cores than WDM \citep[][]{SilkMarsh2013}.

For FDM particle masses $\mfdm \sim 10^{-22} ~ \textup{eV}$, the halo mass function is suppressed for halo masses below $\sim 10^{10} ~ \MSUN$. This eliminates the bulk of halos expected to heat up the IGM and reionize the Universe in the $\Lambda$CDM paradigm. The implied stark difference in the demographics of the first galaxies could potentially be tested by future observations of the Cosmic Dawn and the EoR. The most direct probe of the state of the IGM during those eras is the redshifted 21-cm signal, originating from the hyperfine transition of neutral hydrogen. In principle it has the potential to constrain various details of these epochs such as the exact timing of the reionization, properties of sources responsible for reionizing the IGM, the presence of X-ray sources, the efficiency of various feedback mechanisms, etc. As this signal crucially depend on the population and nature of the sources, it opens a window to both constrain and distinguish between dark matter models. This is what we explore in this article.

A range of existing radio interferometers  such as the Low Frequency Array (LOFAR)\footnote{http://www.lofar.org/} \citep{van13, 2017ApJ...838...65P}, the Precision Array for Probing the Epoch of Reionization (PAPER)\footnote{http://eor.berkeley.edu/} \citep{parsons13}, and the Murchison Widefield Array (MWA)\footnote{http://www.mwatelescope.org/} \citep{bowman13, tingay13} have dedicated substantial resources and efforts to detect fluctuations in the 21-cm signal from the EoR. In parallel, several ongoing experiments such as EDGES \citep{2010Natur.468..796B}, SARAS \citep{2015ApJ...801..138P},  BigHorns  \citep{2015PASA...32....4S}, SciHi  \citep{2014ApJ...782L...9V} and LEDA \citep{2012arXiv1201.1700G} strive to detect the global (i.e.\ sky-averaged) 21-cm signal from the EoR and Cosmic Dawn. Recently, \cite{EDGES2018} presented the first claimed detection of the global 21-cm signal using the EDGES low-band experiment. However, this detection has not been independently confirmed. Furthermore, \cite{Hills2018} argues that better foreground modelling can remove the signal all together.

In anticipation of a confirmed detection of the 21-cm signal, a wide range of theoretical approaches using analytical \citep[e.g.,][]{furlanetto04, 2014MNRAS.442.1470P}, semi-numerical \citep{zahn2007, mesinger07, santos08, choudhury09}, and numerical \citep{Iliev2006,  mellema06, McQuinn2007, shin2008, baek09, Thom09, ghara15a} methods have been considered for modelling this signal, so as to understand the impact of various astrophysical and cosmological processes on this signal.

Reionization in a Universe in which the dark matter is FDM has been studied in terms of the luminosity functions and the production rate of ionizing photons by \cite{Bozek2015} and \cite{Corasaniti2017}. The latter showed that reionization by a redshift $z \simeq 6$, and with a Thomson optical depth value consistent with the Planck results \citep{Planck2015}, is possible in $\Lambda$FDM provided $m_{22} \gtrsim 0.74$ but that constraints from the observed UV luminosity functions for $z=4$ -- 10 imply a lower limit of $m_{22}\geq 1.2$. These authors did not consider the 21-cm signal. \cite{Sarkar2016} did study reionization 21-cm power spectra in FDM cosmologies and derived a lower limit of $m_{22} \gtrsim 0.26$ for reionization to reach 50 percent by $z=8$ and also showed that the resulting 21-cm power is 2 -- 10 higher than for $\Lambda$CDM on observable scales. They did however not consider the X-ray heating and $\lya$ coupling effects on the 21-cm signal. \cite{2018PhRvD..98b3011L} investigated the effect of FDM models on the global 21-cm signal from the Cosmic Dawn and place a lower limit of $m_{22} \ge 50$ for coupling the spin temperature to the gas temperature at $z = 20$, which would be required to explain the timing of the claimed EDGES low-band detection. 

Unlike FDM, the effect of WDM on the 21-cm signal from the Cosmic Dawn and Epoch of reionization has been investigated by many previous studies such as \cite{2014MNRAS.438.2664S} and \cite{2014JCAP...08..007S}. These studies find that WDM delays the emergence of the 21-cm signal and accelerates the impact of X-ray heating.  Recently, \cite{2018ApJ...859L..18S} have used the EDGES low band results to constrain the particle mass of WDM. This study constrains the mass of the WDM particles to be $>3$~keV if the star formation rate at $z = 18$ is dominated by atomic cooling. 

In this article we provide a detailed investigation of the effect of the FDM models on the 21-cm signal in terms of the global signal, the power spectrum as well as  bubble size distributions. In this we take into account the effects of $\lya$ coupling, X-ray heating, as well as photo-ionization. We use a fast analytical framework for this investigation. We also explore the possibilities for ongoing and future 21-cm experiments to constrain the FDM particle mass, $\mFDM$.

The paper is structured as follows. In Section \ref{fdm_details}, we briefly describe the FDM model used in this study as well as their effects on the halo abundance in the Universe. The basic framework of the 21-cm signal from the Cosmic Dawn is presented in Section \ref{sec:model}, and  we present our results in Section  \ref{sec:res}. We summarize our findings in Section \ref{sec:conclude}. Throughout the paper, we adopt the following cosmological parameters: $\Omegam=0.32$, $\OmegaB=0.049$, $\Omega_\Lambda=0.68$, $h=0.67$, $\sigma_8=0.83$, and $n_{\rm s}=0.96$ \citep{Planck2015}. 


\section{Cosmic Dawn in Fuzzy Cosmologies}
\label{fdm_details}

Because the 21-cm signal is strongly governed by the nature of the ionizing and X-ray sources, it is important to have a plausible picture of how galaxy formation would proceed in $\Lambda$FDM. In this section the most important aspects of FDM, especially for galaxy formation, are discussed, some of which will be used in formulating the scenarios to be simulated. In Section~\ref{sec:fuzzy_dm} we use natural units, wherein $\hbar = c = 1$, for convenience, but not elsewhere in this paper.

\subsection{Dynamics \& Background Evolution}
\label{sec:fuzzy_dm}

As discussed by \cite{Marsh2016, Hui2017}, the Lagrangian for the most well-motivated model of FDM --- at least one new ultra-light axion-like scalar field $\phi$ that can appear in the context of string theory --- reads 
\begin{equation}
\mathcal{L} = \frac{1}{2}\partial _{\mu }\phi \partial ^{\mu }\phi - \mfdm^2 F^2 \left [ 1 - \cos \left ( \frac{\phi}{F} \right ) \right ],  
\label{eq:fdm_langrangian}
\end{equation}
where $F$ is a constant predicted to lie somewhere in the range between $\sim 10^{16} ~ \rm GeV$ (the GUT scale) and $\sim 10^{18} ~ \rm GeV$ (the Planck scale). Assuming a Friedmann-Lemaître-Robertson-Walker background, the FDM scalar field is governed by the following Klein-Gordon equation:
\begin{equation}
\ddot{\phi } + 3H\dot{\phi} + \mfdm^2 F \sin \left (\frac{\phi}{F} \right ) = 0,
\end{equation}
where $H = \dot{a}/a$ is the Hubble parameter. The corresponding energy density of the scalar field is
\begin{equation}
\rho_{\textup{FDM}} = \frac{1}{2}\dot{\phi}^2 + \mfdm^2 F^2 \left [ 1 - \cos \left ( \frac{\phi}{F} \right ) \right ].
\label{FDM density}
\end{equation}
When $H \sim t^{-1} \ll \mfdm$, the scalar field starts to oscillate with a decaying amplitude. In this regime the approximate WKB solution of the Klein-Gordon equation is of the form \citep[]{Marsh2016}
\begin{equation}
\phi \simeq \phi_{0} \left( \frac{\aosc}{a} \right)^{3/2}\cos (\mfdm t),
\end{equation}
where $H(\aosc) \sim \tosc^{-1} \sim \mfdm$. From Eq. (\ref{FDM density}) we therefore see that the energy density from then on scales as $\sim a^{-3}$. The time at which this CDM-like behaviour turns on is $\tosc \sim (10^{-22} \mFDM \textup{ eV})^{-1} \simeq 0.2 ~ \mFDM^{-1} \textup{ yr}$, where $\mFDM \equiv \mfdm/10^{-22} \textup{ eV}$. For the particle mass range considered in this paper, this is safely in the radiation-dominated era. In summary, the large-scale evolution of the Universe in $\Lambda$FDM is indistinguishable from $\Lambda$CDM. 

The predicted present-day abundance of FDM is also entirely specified by $\phi_{0}$, $F$, and $\mFDM$ \cite{Diez-Tejedor2017, LymanAlphaConstraints2017}:\footnote{Here we have used the expression for $\Omega_{\rm FDM}$ derived by \cite{LymanAlphaConstraints2017}, but including the logarithmic correction term found by \cite{Diez-Tejedor2017} that becomes important when $\lvert \phi_{0} \lvert/\pi F \simeq 1$ due to anharmonic effects.} 
\begin{equation}
\Omega_{\rm FDM} \simeq 0.16 ~ \mFDM^{1/2} \left( \frac{\phi_{0}}{10^{17} ~ \rm GeV} \right)^2 \ln^{3/2} \left [ \frac{e}{1 - (\phi_{0}/\pi F)^4} \right ].
\end{equation}
As long as $2 \pi F$ is greater than $H_{\rm I}$, the inflationary Hubble scale, then the initial field value of the scalar field, $\phi_{0}$, has a uniform distribution on the interval $[ - \pi F, \pi F]$ \citep[e.g.][]{Hertzberg2008, Marsh2016, Visinelli2018}. The condition $2 \pi F > H_{\rm I}$ will indeed be satisfied for FDM models of interest, because we need $H_{\rm I} \lesssim 4 \times 10^{12} ~ \mFDM^{-1/4} ~ \rm GeV$ in order to evade observational constraints on isocurvature fluctuations \cite{LymanAlphaConstraints2017} --- far smaller than $F > 10^{16} ~ \rm GeV$. We therefore expect that $\phi_{0}^2 \sim F^2$, showing that $F \sim 10^{17} ~ \rm GeV$ (in the range favoured by particle physics) and $\mFDM \sim 1$ (the particle mass motivated by astrophysics) can explain the observed dark matter abundance ($\Omega_{\rm DM} = 0.271$ for our adopted cosmological parameters). Because of this, and in order to maximize the potential of $\Lambda$FDM in solving small-scale problems facing $\Lambda$CDM, we assume throughout the rest of the paper that $\Omega_{\rm FDM} = \Omega_{\rm DM}$.

\subsection{Structure Formation}
\label{sec:Structure_formation}

Modifications to the unfolding of events in $\Lambda$CDM make their first appearance in small-scale structure formation. In the non-relativistic limit and on sub-horizon scales, the amplitude of a density perturbation $\delta_{\textbf{k}}$ with comoving wavenumber $\textbf{k}$ is governed by \citep[e.g.][]{Woo2009, Marsh2016, LymanAlphaConstraints2017}
\begin{equation}
\ddot{\delta}_{\textbf{k}} + 2H\dot{\delta }_{\textbf{k}} = \left ( \frac{3H_{0}^2 \Omegam }{2a^3}  - \frac{\hbar^4 \left | \textbf{k} \right |^4}{4a^{4}\mfdm^2} \right ) \delta_{\textbf{k}},
\label{Eq. FDM perturbation}
\end{equation}
where $H_{0} = 100 ~ h ~ \rm km ~ s^{-1} ~ Mpc^{-1}$ is the Hubble constant. Growth is only possible if the right hand side is positive, which defines a Jeans wavenumber $\kJeans$,
\begin{align}
\kJeans &= \left ( 6H_{0}^2 \Omegam \mfdm^2 a \right )^{1/4} \hbar^{-1} \nonumber \\ 
 &\simeq 36 ~ \mFDM^{1/2} \left ( \frac{ \Omegam h^2}{0.14} \right )^{1/4} \left ( \frac{1 + z}{13} \right )^{-1/4} \textup{ Mpc}^{-1}.
\end{align}
The main implications of this Jeans scale are twofold:
\begin{itemize}
\item There can be no growth of density perturbations and therefore no halo formation on mass scales below $\MJeans = 4\pi (\pi/\kJeans)^3 \overline{\rho}_{\rm m,0}$, where $\overline{\rho}_{\rm m,0} \equiv \overline{\rho}_{\rm m}(z=0)$ is the present-day mean matter density, given by
\begin{equation}
\MJeans \simeq 1.0 \times 10^8 ~ \mFDM^{-3/2} \left ( \frac{ \Omegam h^2}{0.14} \right )^{1/4} \left ( \frac{1 + z}{13} \right )^{3/4} \MSUN.
\label{Jeans mass at z}
\end{equation}
\item In $\Lambda$CDM, density perturbations can only grow significantly in the matter-dominated era. Thus, structure formation will be delayed in $\Lambda$FDM relative to $\Lambda$CDM below mass scales on the order of $\MJeans$ evaluated at the redshift of matter-radiation equality, $z_{\textup{eq}}$:
\begin{equation}
\MJeansEq \simeq 6.5 \times 10^9 ~ \mFDM^{-3/2} \left ( \frac{ \Omegam h^2}{0.14} \right )^{1/4} \left ( \frac{1 + z_{\textup{eq}}}{3400} \right )^{3/4} \MSUN.
\label{JeansMassEquality}
\end{equation}

\end{itemize}
Qualitatively, we therefore expect the halo mass function (HMF) to peak around $\MJeansEq$, and below that drop abruptly so that no halos with masses around $\MJeans$ are present. 

The HMF used in this study needs to capture both of these features in order to accurately model the predicted 21-cm signal. We model the HMF as follows. For the FDM linear power spectrum $\PFDM(k)$, we use the fit from \cite{Hu2000},
\begin{equation}
\PFDM(k) = \TFDM^2(k) \PCDM(k), 
\label{FDM power spectrum}
\end{equation}
with
\begin{equation}
\TFDM(k) \simeq  \frac{\cos(x^3)}{1 + x^8}, 
\label{TFDM}
\end{equation}
in which 
\begin{equation}
x \equiv 1.61 ~ \frac{\mFDM^{1/18} k}{\kJeansEq}.
\end{equation}
We see that there is a significant loss of power relative to $\PCDM(k)$ for $k \gtrsim \kJeansEq$ as expected. With this power spectrum, it is possible to naively compute the HMF using the Press-Schechter or Sheth-Tormen HMF \citep[][]{PressSchechter, ShethTormen2002}, or running a CDM-like N-body simulation \citep[e.g.][]{Schive2016, FirstStarsFDM, Irsic2017}. However, these approaches will not induce a sharp cut-off to the HMF near $\MJeans$ 
since CDM-like N-body simulations and semi-analytical halo mass functions by themselves do not incorporate the relevant pressure-like effect that gives rise to the Jeans scale. \cite{Zhang2017LymanAlpha} followed the evolution of the HMF in $\Lambda$FDM in N-body simulations that both incorporated and ignored the pressure-like effect, showing an expected sharp decline in the low-end HMF when the effect was included.

We model the HMF in a semi-analytic fashion following \cite{SilkMarsh2013}, \cite{Bozek2015}, \cite{MarshHaloModel}, and \cite{Du2017}. These authors modelled the sharp cut-off to the HMF by including a mass dependent barrier $\deltacrit(M)$ for halo formation. This simulates the scale-dependent growth implied by the solution of Eq. (\ref{Eq. FDM perturbation}). \cite{MarshHaloModel} found that $\deltacrit(M)$ could be fitted in a redshift-independent manner,
\begin{align}
\deltacrit(M) ~ &\simeq ~ \GF(M) ~ \deltacritEdS, \nonumber \\ 
\GF(M) ~ &= ~ \hF(x)\exp(a_3 x^{-a_4}) \nonumber \\ &+ ~ [1 - \hF(x)]\exp(a_5 x^{-a_6}), \nonumber \\
x ~ &= ~ M/\MJeans^0, \\
\hF(x) ~ &= ~ \frac{1}{2} \left \{ 1 - \tanh[\MJeans^0 (x - a_2)] \right \}, \nonumber \\
\MJeans^0 ~ &= ~ a_1 \times 10^8 ~ \mFDM^{-3/2} \left ( \frac{ \Omegam h^2}{0.14} \right )^{1/4} h^{-1} \MSUN \nonumber.
\end{align}
Here $\deltacritEdS \simeq 1.686$ is the mass-independent barrier for spherical collapse in the matter-dominated era of $\Lambda$CDM,\footnote{When $1+z \ll (\Omegam / \Omega_\Lambda)^{1/3}$ the barrier asymptotes to $\simeq 1.630$ \citep{TegmarkCDMBarrier}. But during the Cosmic Dawn (and even at $z \simeq 0$), adopting the Einstein-de Sitter value of $\deltacritEdS \simeq 1.686$ is a good approximation.} and the fitted constants are,
\begin{align}
\left \{ a_1, a_2, a_3, a_4, a_5, a_6 \right \} &= \left \{ 3.4,~ 1.0,~ 1.8,~ 0.5,~ 1.7,~ 0.9 \right \}.
\end{align}
Given this expression for $\deltacrit(M)$, we use the Press-Schechter HMF \cite{PressSchechter} with the replacement $\deltacritEdS \rightarrow \deltacrit(M)$,
\begin{equation}
\frac{\partial n (M,z)}{\partial \, \textup{ln}\, M}=\sqrt{\frac{2}{\pi }}\frac{\overline{\rho }_{\rm m,0} }{M}\frac{ \delta _{\textup{crit}}(M) }{\sigma (M,z)}\left | \frac{\partial\, \textup{ln}\, \sigma }{\partial\, \textup{ln}\, M} \right |e^{-\delta _{\textup{crit}}(M)^2 /2\sigma(M,z) ^2}.\
\label{HMF}
\end{equation}
Here $\sigma (M,z)$ is the root-mean-square linear overdensity within a spherical region of comoving mass $M$, computed using the power spectrum in Eq. (\ref{FDM power spectrum}) and normalized so that $\sigma (R = 8~h^{-1}\textup{ Mpc}, z = 0) = \sigma_8$.

\begin{figure}[t]
 \centering
\includegraphics[width=\textwidth]{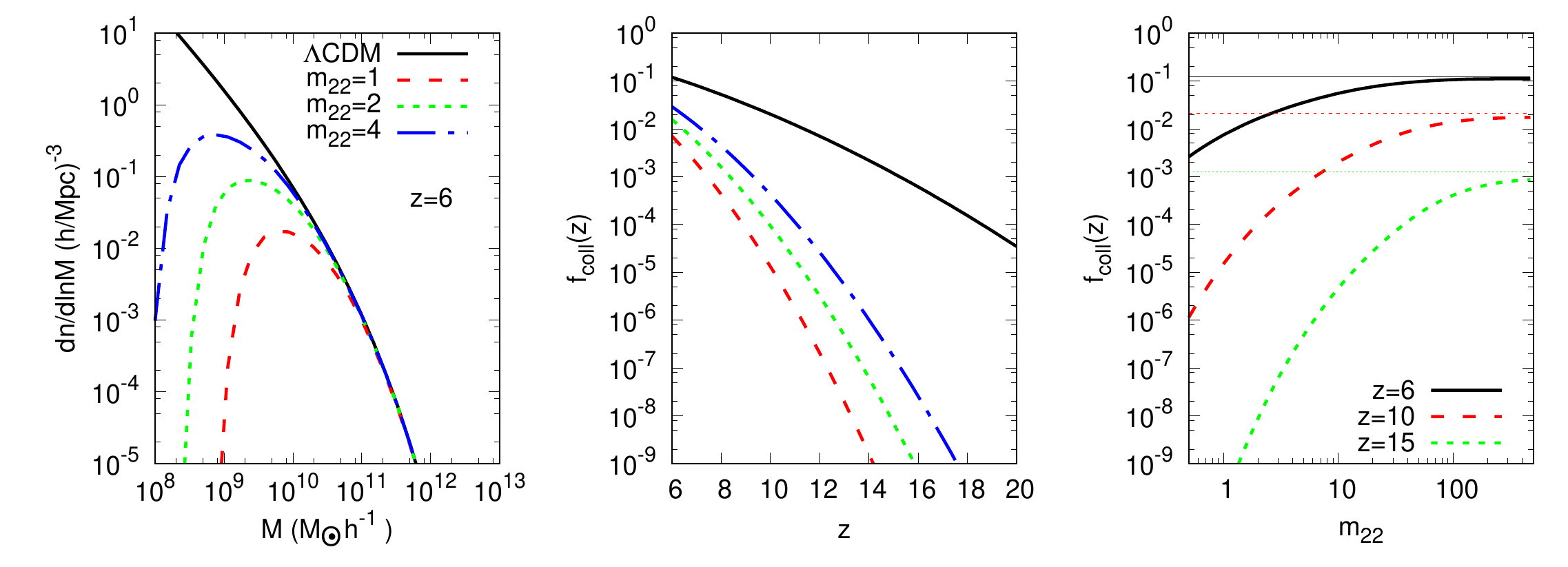}
    \caption{\textbf{Left-hand Panel:} The halo mass functions at redshift 6 for $\Lambda$CDM, and $\Lambda$FDM with a selected few values of $\mFDM$ considered in this work. \textbf{Middle Panel:} Redshift evolution of the collapse fractions corresponding to the same scenarios as in the left-hand panel. \textbf{Right-hand Panel:} Dependence of the collapsed fraction on the FDM particle mass, $m_{22}$ at different redshifts. The horizontal lines mark the collapse fractions corresponding to $\Lambda$CDM. The minimum Virial temperature for star formation is fixed to $T_{\rm vir} = 10^4 ~ \rm K$.  While $T_{\rm vir}$ determines the minimum halo mass for star formation in $\Lambda$CDM, $M_{\rm min}$ corresponding to $\Lambda$FDM models is estimated following Eq. \ref{Mmin}. }
   \label{image_p6dndmfcoll}
\end{figure}

The simple prescription $\deltacritEdS \rightarrow \deltacrit(M)$ was first used by \cite{SilkMarsh2013} and \cite{Bozek2015}. The HMF in Eq. (\ref{HMF}) is plotted at $z = 6$ for different $\mFDM$ in the left-hand panel of Figure \ref{image_p6dndmfcoll}. It is seen that the HMF for FDM peaks at $\sim 4 \times 10^9 ~h^{-1}~\MSUN$ for $\mFDM = 1$, fairly close to what we expected from Eq. (\ref{JeansMassEquality}). A dramatic deviation from $\Lambda$CDM is evident for halo masses below this peak. For example, with $m_{22}=1$ the number density of halos with masses $\sim 10^9 ~\MSUN$ is four dex below the predicted number density in $\Lambda$CDM. 

The fact that Eq. (\ref{HMF}) captures the peak and the sharp decline in the HMF is an improvement over the fit provided by \cite{Schive2016}, who ran N-body simulations with the FDM power spectrum from Eq. (\ref{FDM power spectrum}), but ignored the pressure-like effect on small scales. However, Eq. (\ref{HMF}) is not self-consistent. This is because a proper derivation of the HMF for a mass-dependent barrier should make use of the excursion set formalism \citep{Bond1991} and not the mere replacement $\deltacritEdS \rightarrow \deltacrit(M)$.

The excursion set problem for $\Lambda$FDM was numerically solved by \cite{Du2017}, and compared with the approach we adopt here. These authors found that a self-consistent solution of the excursion set problem will yield a slightly sharper cut-off in the HMF, resulting in even fewer low-mass halos than predicted by Eq. (\ref{HMF}). We neglect this for the following reasons:
\begin{itemize}
\item \cite{Du2017} found that the difference between the excursion set solution and the adopted HMF here is minimized at high redshifts relevant for the Cosmic Dawn.

\item In order to draw strong conclusions from the predicted evolution of the 21-cm signal, we take a conservative approach. A sharper cut-off in the HMF would only exacerbate the deviation from $\Lambda$CDM, and therefore make our conclusions even stronger.

\end{itemize}

To conclude this section we would like to point out that there exist `extreme' versions of FDM, having an initial field displacement $\lvert \phi_{0} \lvert/\pi F \simeq 1$ \cite{Zhang2017Extreme, Schive2018_HaloAbundance, Leong2018}. In this case, anharmonic effects from the potential in Eq. \ref{eq:fdm_langrangian} can, for a given $\mFDM$, lead to a far greater number of low-mass halos than the `vanilla' version of FDM described above. As the properties of the HMF determine the galaxy population in our models, as explained below, the presence of large numbers of low-mass halos would lead to very different results for our calculations. Indeed, \cite{Leong2018} finds that $\Lambda$FDM can evade Ly$\alpha$ constraints at least as easily as $\Lambda$CDM if $\delta \theta_{0} \equiv \lvert \phi_{0} \lvert/F - \pi  ~ \lesssim 0.087$, with a best fit somewhere in the range $0.044 \lesssim \delta \theta_{0} \lesssim 0.061$ for $\mFDM = 1.1$. However, the prior probability of the initial field value being this close to the top of the potential is at most $P(\delta \theta_{0} < 0.087) = 0.087/\pi \simeq 0.028$, or in other words the `extreme' FDM model requires improbable initial conditions.\footnote{Since a small value of $\delta \theta_{0}$ only increases the abundance of low-mass halos hosting relatively few stars, it seems unlikely that anthropic arguments can escape this conclusion.} Given this, we will only consider `vanilla' FDM in this paper.

\subsection{Galaxy Formation}

The suppression of structure formation on small scales also implies that galaxy formation should be significantly delayed in $\Lambda$FDM relative to $\Lambda$CDM. More specifically, since structure formation is only suppressed below $\sim \MJeansEq$ in $\Lambda$FDM, the hierarchical structure formation above this scale implies that the first galaxies form in halos with masses $\sim \MJeansEq$. For comparison, $\Lambda$CDM predict that the first proper galaxies formed in atomic cooling halos with characteristic masses of $\sim 10^{8} ~\MSUN$ set by efficient $\lya$ cooling \citep[e.g.][]{Greif2008}. These would have formed relatively early compared to the $\sim \MJeansEq$ halos in $\Lambda$FDM since $\MJeansEq \gtrsim 10^{8} ~\MSUN$ for $\mFDM \lesssim 16$.

Furthermore, with no early minihalos with mass $M \sim 10^{6} ~\MSUN$ in which $\textup{H}_{2}$ cooling is possible \citep[see e.g.][]{Tegmark1997, Bromm2013}, $\Lambda$FDM predicts that the sites for Pop III star formation would be the first atomic cooling halos with masses $\sim \MJeansEq$. The expectation of delayed galaxy formation and radically different host halos for Pop III star formation in $\Lambda$FDM was recently confirmed in the hydrodynamical simulations of \cite{FirstStarsFDM}. 

The process of reionization as well as the thermal state of the IGM will be closely related to $f_{\rm coll}$, the fraction of baryons in collapsed structures wherein star formation can proceed. If star formation is possible in halos above a minimum mass $M_{\rm min}$, then the collapse fraction can be computed by integrating the HMF,
\begin{equation}
f_{\rm coll}(z)=\frac{1}{\bar{\rho}_{\rm m,0}}\int_{M_{\rm min}}^{\infty} {\rm d} M ~\frac{\partial n(M,z)}{\partial \, \textup{ln}\, M}.
\label{equ_fcoll1}
\end{equation}
In $\Lambda$FDM, both the inability of virialized gas to cool efficiently and the stability against gravitational collapse below the Jeans scale compete to determine $M_{\rm min}$. As remarked earlier, minihalos where $\textup{H}_{2}$ cooling is possible can be completely neglected in $\Lambda$FDM. Thus, efficient cooling is only possible in halos with virial temperatures above $T_{\rm vir} \simeq 10^4$ K, determined by the onset of rapid Ly$\alpha$ cooling. The halo mass corresponding to this limit is \citep[e.g.][]{LoebFurlanettoFirstGalaxies}
\begin{align}
M_{\rm Ly\alpha} &\simeq ~ 6.3 \times 10^7 ~ \left ( \frac{T_{\rm vir}}{10^4 ~ \rm K} \right )^{3/2} \left ( \frac{ \Omegam h^2}{0.14} \right )^{-1/2} \left ( \frac{\deltavir}{18\pi^2} \right )^{-1/2} \nonumber \\ &\times ~ \left ( \frac{\mu}{0.6} \right )^{-3/2} \left ( \frac{1 + z}{13} \right )^{-3/2} \MSUN,
\label{Lyman alpha cooling}
\end{align} 
where $\deltavir \equiv \rho_{\rm vir}/\bar{\rho}_{\rm m}$ is the mean overdensity of a virialized halo, and $\mu$ is the mean molecular weight of the baryons at temperature $T_{\rm vir}$. Since we are focusing on the high-redshift matter-dominated Universe, we adopt $\deltavir = 18 \pi^2$ \cite{1998ApJ...495...80B, Tegmark2006}. The mean molecular weight is $\mu \simeq 1.2$ for neutral primordial gas, and $\mu \simeq 0.6$ for ionized primordial gas. Even though hydrogen in collisional ionization equilibrium at a temperature $< 2 \times 10^4 ~ \rm K$ is mostly neutral, we adopt $\mu = 0.6$, which crudely takes into account photoionization of the accreting gas from the IGM. Given this, the minimum halo mass for star formation will simply be,
\begin{equation}
M_{\rm min} = \textup{max} \left( ~ M_{\rm Ly\alpha} ~ , ~ \MJeans  ~ \right ).
\label{Mmin}
\end{equation}
Comparing Eq. (\ref{Lyman alpha cooling}) and Eq. (\ref{Jeans mass at z}), we see that $\MJeans > M_{\rm Ly\alpha}$ for redshifts $1+z > 11 ~ \mFDM^{2/3}$. However, both $\MJeans$ and $M_{\rm Ly\alpha}$ lie at masses below the peak in the HMF, and therefore do not have much of an effect on the collapse fraction $f_{\rm coll}(z)$. The collapse fractions for $\Lambda$CDM and $\Lambda$FDM are plotted in the middle panel of Figure \ref{image_p6dndmfcoll}. The collapse fraction for $\Lambda$FDM, with $\mFDM \sim \mathcal{O}(1)$, remains significantly smaller than $\Lambda$CDM even towards the end of the EoR at $z \simeq 6$. Furthermore, the redshift gradient of $f_{\rm coll}(z)$ is seen to be steeper for $\Lambda$FDM. From these observations, a later and more rapid heating and reionization of the IGM is expected for $\Lambda$FDM. The right-hand panel of the Figure \ref{image_p6dndmfcoll} show the dependence of the collapsed fraction on $m_{22}$ parameter at different redshifts. Although $f_{\rm coll}$ decreases drastically with $m_{22}$, it approaches the $\Lambda$CDM values for large $m_{22}$ values. Even for $\mFDM = 15$ we see that the collapse fractions for $\Lambda$FDM and $\Lambda$CDM differ by a factor of $\sim 100$ at $z = 15$.  


\section{Model of the 21-cm signal}
\label{sec:model}
\subsection{Analytical Model of Cosmic Dawn and EoR}

The analytical model used in this work follows previous works such as \cite{Pritchard07, 2005ApJ...630..643M}. The model incorporates the effects of both UV and X-ray photons and tracks the evolution of the volume averaged ionization fractions $x_i$ and $x_e$ which correspond to the highly ionized \HII ~regions and largely neutral gas in the IGM outside these regions. It also estimates the kinetic temperature ($T_K$) of the largely neutral medium outside the \HII ~regions, while $T_K$ is assumed to be $\sim 10^4$ K in the \HII ~regions.  The basic formalism of the model is described as below.

The rate of production of the UV photons per baryon can be expressed as,
\begin{equation}
\Lambda_i = \zeta \frac{ {\rm d} f_{\rm coll}}{ {\rm d} t},
\end{equation}
where $\zeta \equiv N_{\rm ion} f_\star f_{\rm esc}$ is the ionization efficiency parameter. The ionization state of the IGM therefore crucially depends on quantities such as the star formation efficiency ($f_\star$), escape fraction of the UV photons ($f_{\rm esc}$), and mean number of ionizing photons produced per stellar baryon ($N_{\rm ion}$). We describe how we set these three quantities in Sections~\ref{sec:scen} and \ref{sec:source}. 

The ionization and thermal state of the mostly neutral medium beyond these highly ionized \HII ~regions crucially depend on the X-rays produced by these sources. The emissivity of the X-ray photons from the sources is assumed to follow the star formation rate density. We use two X-ray parameters, namely $f_X$ and $\alpha_X$ which quantify the spectral distribution which is modelled as,
\begin{eqnarray}
\epsilon_X(\nu) = \frac{L_0}{h\nu_0}\left(\frac{\nu}{\nu_0}\right)^{-\alpha_X-1},
\end{eqnarray}
where $L_0=1\times 10^{40} f_X ~\rm erg ~s^{-1} ~Mpc^{-3}$, $h\nu_0 = 1 ~\rm keV$.

The 21-cm signal from the cosmic dawn depends also on the $\lya$ photon flux from  the sources. Here we estimate the average $\lya$ photon flux using the formalism of \cite{2006MNRAS.372.1093F}. We consider the stellar emission and the X-ray excitation of the neutral hydrogen to estimate the average $\lya$ background. The details of the source model used in this study is described in the following section.

\subsection{The star formation efficiency}
\label{sec:scen} 

\begin{figure}
\begin{center}
\includegraphics[width=\textwidth]{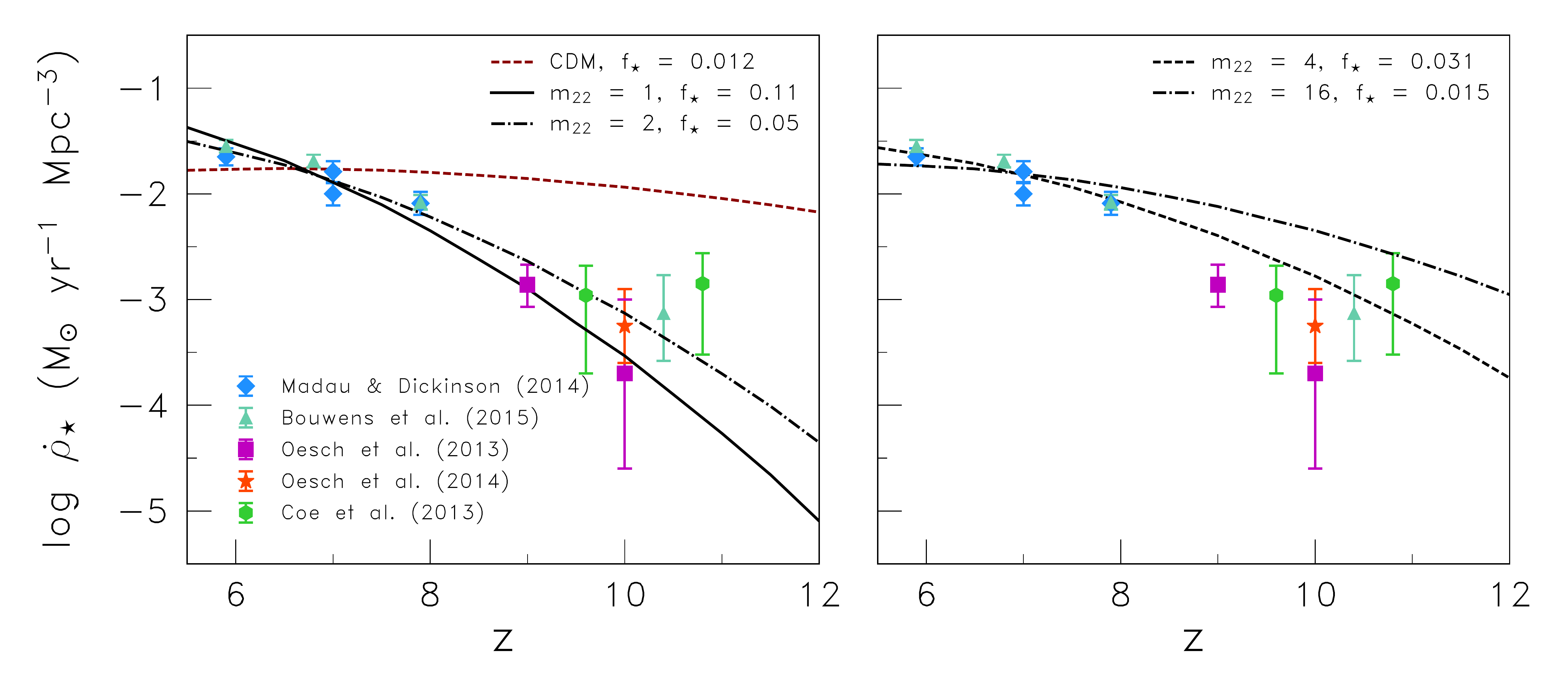}
   \caption{Predicted cosmic star formation history for the main scenarios considered in this work at redshifts between $z = 5.5$ and $z = 12.0$, along with observational constraints. The star formation efficiency $f_{\star}$ is tuned individually for each dark matter scenario in order to roughly fit the observational constraints on the star formation history. In both panels we plot observational constraints compiled by \cite{MadauDickinson2014} from $z = 5.9-7.9$, as well as constraints at higher redshifts $z \sim 9 - 11$ from \cite{Coe2013}, \cite{Oesch2013}, \cite{Oesch2014}, and \cite{Bouwens2015}. The limits at redshifts $z \sim 9 - 11$ should be seen as lower limits to $\dot{\rho}_{\star}$ as they only correspond to galaxies brighter than $M_{\rm UV} \sim -17$. \textbf{Left-hand panel:} The evolution of the cosmic star formation rate in $\Lambda$CDM and $\Lambda$FDM with $\mFDM = 1, 2$.  \textbf{Right-hand panel:} Same as the left panel but for $\Lambda$FDM with $\mFDM = 4, 16$. }
  \label{StarFormationHistory}
\end{center}
\end{figure}

Only a fraction, denoted $f_{\star}$, of the baryons in an atomic-cooling halo will be able to form stars. Following previous work we assume for simplicity that this fraction is independent of the halo mass and redshift \citep[e.g.][]{FurlanettoGlobal2006, Pritchard07, Mesinger2013}. It then follows that the cosmic star formation rate per unit comoving volume is,
\begin{equation}
\frac{{\rm d} \rho_{\star}}{{\rm d} t} = f_{\star} \frac{\OmegaB}{\Omegam} \overline{\rho }_{\rm m,0} \frac{ {\rm d} f_{\rm coll}}{ {\rm d} t}.
\label{Cosmic star formation rate}
\end{equation}
We can then tune $f_{\star}$ to roughly be consistent with observational constraints on the cosmic star formation rate for redshifts $z \gtrsim 6$. In Figure \ref{StarFormationHistory} we compare the theoretically derived $\dot{\rho}_{\star}$ from Eq. (\ref{Cosmic star formation rate}) with observational constraints on the cosmic star formation rate. For redshifts $z = 5.9 - 7.9$ we use data compiled by \cite{MadauDickinson2014}. At higher redshifts of $z \sim 9 - 11$ we also show observational constraints from \cite{Coe2013}, \cite{Oesch2013}, \cite{Oesch2014}, and \cite{Bouwens2015}. These constraints at higher redshifts should be seen as lower limits to $\dot{\rho}_{\star}$ as the UV luminosity function is only integrated down to $M_{\rm UV} \simeq -17.7$ for the data points by \cite{Coe2013}, \cite{Oesch2013}, and \cite{Oesch2014}, whereas the constraint at $z \simeq 10.4$ by \cite{Bouwens2015} integrates down to $M_{\rm UV} \simeq -17.0$. 

We choose $f_{\star}$ mainly so that the tighter constraints at $z \sim 6 - 8$ are approximately respected. Doing this, it is seen that $\Lambda$FDM with $\mFDM = 1$ is possibly in tension with current lower limits to $\dot{\rho}_{\star}$ at $z \sim 10 - 11$, whereas $\mFDM = 2$ is not. This is broadly consistent with the $2\sigma$ lower bounds $\mFDM \geq 1.2$ and $\mFDM \geq 1.6$ derived by a detailed modelling of the UV luminosity function for $\Lambda$FDM by \cite{Schive2016} and \cite{Corasaniti2017} respectively. The values of $f_{\star}$ needed to be consistent with data is seen to increase with $\mFDM$ from $f_{\star}(\mFDM = 1) \simeq 0.11$ to $f_{\star}(\mFDM = 16) \simeq 0.015$. For $\Lambda$CDM, which corresponds to the limit $\mFDM \rightarrow \infty$, we find $f_{\star}(\mFDM \rightarrow \infty) \simeq 0.012$. The resulting $f_{\star}$-$\mFDM$ relation can be fitted with the following expression.
\begin{equation} 
f_{\star, \rm fit}(\mFDM) = 0.012 + 0.098 \exp[ - 0.95(\mFDM - 1)^{1/2}].
\label{fit to fstar}
\end{equation}
This fit agrees exactly with our chosen values for $f_{\star}(\mFDM = 1)$ and $f_{\star}(\mFDM \rightarrow \infty)$, and in between only deviates by less than $3.5\%$. The fit $f_{\star, \rm fit}(\mFDM)$ will only be used in our study of the parameter space in Section \ref{sec:param} where we need a continuous relationship between $f_{\star}$ and $\mFDM$. 

The fact that we need a higher $f_{\star}$ for lower $\mFDM$ simply reflects the fact that $\Lambda$FDM produces fewer low-mass halos than $\Lambda$CDM which means that in order to not violate constraints on $\dot{\rho}_{\star}$ the average fraction of baryons ending up in stars needs to be larger. Beyond this empirical constraint, such a $f_{\star}$-$\mFDM$ relationship could also plausibly arise naturally. Observations indicate that $f_{\star}$ is a non-linear function of the halo mass --- growing with halo mass for $M \lesssim 10^{12} ~ \MSUN$ \citep[e.g.][]{Behroozi2013,Puebla2017}. Since lower values of $\mFDM$ suppresses the formation of low-mass halos, the halo mass-weighted average of $f_{\star}$ would decrease with $\mFDM$.

Figure \ref{StarFormationHistory} also shows the chosen values of $f_{\star}$ for our four FDM models which will be describe latter in section \ref{sec:mainscen}. We also list them in Table \ref{tab1} . For the parameter study we adopt the fit in Eq. (\ref{fit to fstar}) which approximately reproduces the $f_{\star}$-$\mFDM$ relation in the main scenarios.

\subsection{Source properties}
\label{sec:source}

The ionization and thermal state of  the IGM during the Cosmic Dawn and the EoR crucially depend on the properties of the radiating sources present during these epochs.   However, there is a huge uncertainty regarding these. For the 21-cm signal, the most important photons are UV, X-rays and $\lya$ photons which are drivers of ionization, heating and $\lya$ coupling respectively. 
\begin{itemize}
\item \textbf{Ionizing photons}: The relative contributions from the PopIII and PopII stars to the total ionizing photon budget is not well understood. We fix $\NionPopII = 4000$ and $\NionPopIII = 30000$ for the PopII and PopIII source model respectively.  The escape fraction of ionizing photons $f_{\rm esc}$ is taken to be 0.1. For a given fraction $\fPopII$ of stars that are PopII, the total ionization budget will then be
\begin{equation}
\Nion = \fPopII \NionPopII + (1 - \fPopII) \NionPopIII.
\label{equ_fpopii}
\end{equation}
The parameters $\Nion$ and $\fPopII$ are determined as follows:
\begin{enumerate}
\item First we estimate the approximate value of $\Nion$ needed to reionize the Universe by $z = 6$ for a given value of $\mFDM$.
\\
\item Next, if $ \NionPopIII \geq \Nion \geq \NionPopII$ using Eq. \ref{equ_fpopii} we determine the corresponding value of $\fPopII$. If $0 \leqslant \fPopII \leqslant 1$, we adopt this value.
\\
\item If $\Nion > \NionPopIII$, our models cannot reionize the Universe by $z = 6$. We do not consider these cases here. 
\\
\item If $\Nion < \NionPopII$, the Universe can easily be reionized by $z = 6$ from PopII stars alone, so we set $\fPopII = 1$. 
\end{enumerate}

\item \textbf{X-rays}: Similar to the ionizing sources of reionization, the X-ray sources are also uncertain. Mini-quasars, supernova remnants, X-ray binaries and the hot interstellar medium in starburst galaxies are some of the possible X-ray sources during this epoch. We choose $f_X = 1$ and $\alpha_X = 0.5$ as our fiducial X-ray spectrum which corresponds to a mini-quasar type X-ray source. In the parameter study in Sect.~\ref{sec:param} we consider a wide range of values for $f_X$ and $\alpha_X$.

\item \textbf{\boldmath$\lya$ photons}: We follow \cite{2006MNRAS.367.1057P} to model the $\lya$ emission from the sources. We assume a power law spectrum  $\epsilon_s(\nu) \propto \nu^{-\alpha_{s}-1}$ between $\lya$ and $\lyb$ and between $\lyb$ and the Lyman limit, where the power law indices can differ. The spectral index $\alpha_s$ between $\lya$ and $\lyb$ is taken to be 0.14 and 1.29  for PopII and PopIII stars, respectively. For PopII stars, the spectrum is normalized such that the number of $\lya$ photons per baryon in the range $\lya$-$\lyb$ is 6520 and we adjust the spectral index in the range $\lya$-Lyman limit so that the total number of photons per baryon for this wavelength regime is  9690. These numbers are 2670 and 4800 respectively for PopIII stars. Note that we consider $\lya$ photon contributions from both PopII and PopIII stars in models with $0<\fPopII<1$.
\end{itemize}
Note that the use of a redshift-independent fraction of PopII stars (i.e. $\fPopII = \rm constant$) is a consequence of assuming a redshift-independent escape fraction $f_{\rm esc}$ and star formation efficiency $f_{\star}$. In reality, the fraction of PopII and PopIII stars would evolve with time as halos and the IGM become enriched with metals. The exact conditions needed to transition from massive PopIII star formation to the formation of relatively low-mass PopII stars are still debated. Ignoring the presence of dust, CII or OI mediated cooling can probably induce fragmentation to low-mass PopII stars for $\mathrm{[C/H]} \gtrsim - 3.5$ or  $\mathrm{[O/H]} \gtrsim - 3.0$ respectively \cite{BrommLoeb2003}. However, dust cooling could push the critical metallicity for low-mass PopII star formation down to $-6 \lesssim  \log{Z/Z_{\odot}} \lesssim -5$ \citep[e.g.][]{Schneider2006, Dopcke2011}, but it may also be the case that the dust is evacuated before this can happen \cite{Fukushima2018}. 
These considerations imply that a realistic modelling of the evolution of the PopII fraction in the context of the analytical models used here would be fraught with uncertainties. However, $\fPopII = \rm constant$ may be a good approximation for $\Lambda$FDM given the suppression of small-scale structure formation. In particular, low-mass halos in $\Lambda$FDM would have few, if any, progenitors of lower masses, indicating a monolithic collapse. With few progenitors containing enriched gas, the transition from PopIII to PopII star formation would be prolonged, and so $\fPopII$ would remain almost constant. This is consistent with the hydrodynamical simulations of \cite{FirstStarsFDM}, who found that PopIII star formation persists down to $z \sim 7$ for $\mFDM = 1$.

\subsection{\HII~ Bubble Size Distribution}
\label{sec:BSD}

To correctly model the fluctuations in the 21-cm signal towards the end of the EoR, we need to have an appropriate model of the size distribution of \HII~ regions around ionizing sources. We model the \HII~ bubble size distribution (BSD) mainly following the formalism of \cite{2004ApJ...613....1F} and \cite{2005ApJ...630..643M} with some modifications for $\Lambda$FDM. In the formalism of \cite{2004ApJ...613....1F} we consider a region of mass $M$ with linear overdensity $\delta_{\rm M}$, and RMS linear overdensity $\sigma(M) \equiv \sigma(M,z=0)$. The IGM within this region of mass $M$ will be ionized at a redshift $z$ if there are a sufficient number of ionizing photons at that time and in that region. Mathematically this criterion can be written as
\begin{equation}
\zeta ~ f_{\rm coll}[~ z,~ \delta_{\rm M},~ \sigma(M)~] \geq 1.
\label{HII bubble criterion}
\end{equation}
If this is fulfilled, there is at least one ionizing photon per hydrogen atom in our region of mass $M$. The collapse fraction in this expression can be derived using the extended Press-Schechter formalism \citep{Bond1991}. In $\Lambda$CDM, the barrier for halo formation $\deltacritEdS$ is constant and thus there is an analytical solution for the collapsed fraction which can be expressed as
\begin{equation}
f_{\rm coll}\left[~ z,~ \delta_{\rm M},~ \sigma(M)~\right] = \textup{erfc} \left [ \frac{\deltacritEdS(z) - \delta_{\rm M}}{\sqrt[]{2\left [\sigma^2_{\rm min} - \sigma^2(M) \right ]}} \right ].
\label{conditional collapse fraction}
\end{equation}
In this notation, all the redshift dependence from the growth of linear density perturbations has been absorbed into the barrier: $\deltacritEdS(z) = \deltacritEdS/D(z)$ where $D(z)$ is the growth factor of linear density perturbations. Because of this, the growth factor is not incorporated into $\sigma_{\rm min} = \sigma(M_{\rm min})$ even though $M_{\rm min}$, given by Eq. (\ref{Mmin}), is redshift dependent.

Let $\delta_{\rm M} = \delta_{\rm x}(M,z)$ be defined such that 
\begin{equation}
\zeta ~ f_{\rm coll}\left[~ z,~ \delta_{\rm x}(M,z),~ \sigma(M)~\right] = 1.
\label{HII bubble criterion2}
\end{equation}
The physical interpretation of the barrier $\delta_{\rm x}(M,z)$ is that regions with linear overdensity $\delta_{\rm M} \geq \delta_{\rm x}(M,z)$ will be ionized. For $\Lambda$CDM we can use Eq. (\ref{conditional collapse fraction}) to invert Eq. (\ref{HII bubble criterion}) and obtain an analytical expression for $\delta_{\rm x}(M,z)$. However, for $\Lambda$FDM the barrier for halo formation is not constant and Eq.~(\ref{conditional collapse fraction}) can therefore not be used to find $\delta_{\rm x}(M,z)$. However, an excellent approximate solution for $\delta_{\rm x}(M,z)$ can be found as follows:

\begin{itemize}
\item We can always find some value of the RMS linear overdensity $\sigma_{\rm min, (1)}$ such that:
\begin{equation}
\textup{erfc} \left [ \frac{\deltacritEdS(z)}{\sqrt[]{2} ~ \sigma_{\rm min, (1)}} \right ] = f_{\rm coll}(z),
\label{Def of sigma_min1}
\end{equation}
where $f_{\rm coll}(z)$ is the global collapse fraction in $\Lambda$FDM given by Eq. (\ref{equ_fcoll1}). We can readily invert Eq. (\ref{Def of sigma_min1}) to find,
\begin{equation}
\sigma_{\rm min, (1)} = \frac{\deltacritEdS(z)}{\sqrt[]{2} ~ \textup{erf}^{-1} \left [ 1 - f_{\rm coll}(z) \right ]}.
\end{equation}
In the limit when $\mFDM \rightarrow \infty$ we get $\sigma_{\rm min, (1)} \rightarrow \sigma_{\rm min}$. Thus, $\sigma_{\rm min, (1)}$ yields 1$^{\rm st}$-order corrections to the erfc-formula for the conditional collapse fraction in $\Lambda$FDM:
\begin{equation}
f_{\rm coll}^{(1)}[~ z,~ \delta_{\rm M},~ \sigma(M)~] = \textup{erfc} \left [ \frac{\deltacritEdS(z) - \delta_{\rm M}}{\sqrt[]{2\left [\sigma^2_{\rm min, (1)} - \sigma^2(M) \right ]}} \right ].
\label{FDM conditional collapse}
\end{equation}
\item Next, we can use Eq. (\ref{FDM conditional collapse}) and Eq. (\ref{HII bubble criterion}) to derive $\delta_{\rm x}^{(1)}(M,z)$, the 1$^{\rm st}$-order approximation to $\delta_{\rm x}(M,z)$:
\begin{equation}
\delta_{\rm x}^{(1)}(M,z) = \deltacritEdS(z)~ - ~ \sqrt[]{2}~K(\zeta)~\sqrt[]{\sigma^2_{\rm min, (1)} - \sigma^2(M)},
\label{FDM bubble barrier}
\end{equation}
where $K(\zeta) = \rm{erf}^{-1}(1 - \zeta^{-1})$.
\end{itemize}

Using the approximate barrier for \HII~ bubble formation in Eq. (\ref{FDM bubble barrier}), the \HII~ bubble mass distribution becomes:
\begin{equation}
\frac{\partial n_{\rm b} (M,z)}{\partial \, \textup{ln}\, M}=\sqrt{\frac{2}{\pi }}\frac{\overline{\rho }_{\rm m,0} }{M}\frac{ \left | T(M,z) \right |}{\sigma (M)}\left | \frac{\partial\, \textup{ln}\, \sigma }{\partial\, \textup{ln}\, M} \right |e^{-\delta_{\rm x}^{(1)}(M,z)^2 /2\sigma(M) ^2},\
\label{HII Bubble Mass Function}
\end{equation}
where,
\begin{equation}
T(M,z) = \sum\limits_{n=0}^5 \frac{(-S)^n}{n!}\frac{\partial^n \delta_{\rm x}^{(1)}(M,z)}{\partial S^n}, ~~~ S \equiv \sigma(M)^2.
\end{equation}
The function $T(M,z)$ is a fit that takes into account the non-linearity of the \HII~ bubble formation barrier in Eq. (\ref{FDM bubble barrier}). If Eq. (\ref{FDM bubble barrier}) is expanded linearly in $S = \sigma(M)^2$, we recover the exact same bubble mass distribution formula as in \cite{2004ApJ...613....1F}. We do not perform such an expansion because the non-linearity in the bubble formation barrier is not negligible in $\Lambda$FDM. The fitting function $T(M,z)$ was found by \cite{ShethTormen2002}, and yields an approximate mass function when the barrier is slightly non-linear in $S$. \cite{GeneralBarrier} confirmed that the \HII~ bubble mass function computed using $T(M,z)$ and the form of the barrier function in Eq. (\ref{FDM bubble barrier}) yielded a good approximation to the bubble mass function derived numerically in the extended Press-Schechter formalism.

\subsection{21-cm Signal}
\label{21cm}

The differential brightness temperature of the 21-cm signal at a region with coordinate $\mathbf{x}$  can be written as,
\begin{equation}
 \TB (\mathbf{x}, z) =  27 ~ x_{\rm HI} (\mathbf{x}, z) [1+\delta_{\rm B}(\mathbf{x}, z)] \left(\frac{\OmegaB h^2}{0.023}\right) \times \left(\frac{0.15}{\Omegam h^2}\frac{1+z}{10}\right)^{1/2}  \left(1-\frac{\TCMB}{\TS} \right)\,\rm{mK},
\label{eq_tb}
\end{equation}
where $x_{\rm HI}$, $\delta_{\rm B}$, $\TCMB=2.73 \times (1+z)$ K and $\TS$ denote the neutral fraction, density contrast, the CMBR temperature and the spin temperature of hydrogen gas at position  $\mathbf{x}$ at redshift $z$. Note that the above expression of $\TB$ does not include the contribution from the peculiar velocities of the gas in the IGM. As the effect of peculiar velocities is not very significant in presence of the spin temperature fluctuations \cite{ghara15a, ghara15b}, we will ignore their contribution to the brightness temperature fluctuations in this study.

The fluctuations in the differential brightness temperature can, to first order, be linearly expanded as \citep[see e.g.][]{Pritchard07}
\begin{equation}
\deltaTb = \beta_{\rm B} \deltaB + \betaX \deltaX + \betaAlpha \deltaAlpha + \betaT \deltaT,
\label{eq:fluc}
\end{equation}
where $\deltaB$ is the baryon density fluctuation, $\deltaX$ is the fluctuation in the neutral
fraction, $\deltaAlpha$ the fluctuation in $\xAlpha$, and $\deltaT$ the fluctuation in the kinetic temperature $\TK$. The expression of the $\beta$ coefficients can be found in \cite{Pritchard07}. The spherically averaged power spectrum of $\deltaTb$ can then be written as 
\begin{equation}
P_{\deltaTb}(k) = {\overline{\deltaTb}}^2 P_{\delta \delta}(k) ({\beta^\everymodeprime}^2 + {2\beta^\everymodeprime}/3 + 1/5),
\label{eq:pstb}
\end{equation}
where $P_{\delta \delta}(k)$ and $\overline{\deltaTb}$ denote the power spectrum of the density field and the mean brightness temperature respectively. The quantity $\beta^\everymodeprime$ can be expressed as
\begin{equation}
\beta^\everymodeprime = \beta_{\rm B} - \betaX \bar{x}_e g_e/(1+\bar{x}_e) + \betaT g_T + \betaAlpha W_\alpha,
\label{eq:beta}
\end{equation}
where we assume $\deltaT=g_T(k,z)\delta, ~\deltaAlpha=W_\alpha(k,z) \delta$ and $\delta_e=g_e\delta $ with $\delta_e=(1-1/x_e)\deltaX$. We will present our results in terms of the dimensionless power spectrum $\Delta^2(k) = k^3 P_{\deltaTb}(k)/2\pi^2$.  We note that the linear approximation for the 21-cm power spectrum is a relatively crude one as the cross-terms and non-linear terms can contribute substantially, see \cite{2007ApJ...659..865L} and \cite{2012MNRAS.422..926M}.


\section{Results}
\label{sec:res}

The main motive of this paper is to study the impact of different FDM models on the cosmological 21-cm signal from the EoR and CD in terms of different quantities such as the power spectrum of the brightness temperature, global signal, bubble statistics, etc. which can provide useful insights on these epochs. To gain intuition we first study in Sect.~\ref{sec:mainscen} a set of four FDM models which differ in their value of $m_{22}$ and compare them to a CDM model. For this limited set of models we can present both the global signal evolution and the power spectra in detail. After that we present in Sect.~\ref{sec:param} a parameter study where we vary both $m_{22}$ and the X-ray parameters $f_X$ and $\alpha_X$.

\begin{table}
\begin{center}
\centering
\small
\tabcolsep 3pt
\renewcommand\arraystretch{1.5}
   \begin{tabular}{c c c c c c c c c c c}
\hline
\hline
    Scenario  & DM model  & $m_{22}$ & $f_{\star}$	& $N_{\rm ion}$ & $f_{\rm PopII}$ & $\tau$ ($10^{-2}$) & $<\TB>_{\rm min}$ (mK) &  $z_{\TB,\rm min}$ & $\Delta^2_{\rm max} ~ ^{\rm a}$ & $z_{\Delta_{\rm max}}$  \\

\hline
\hline

   S0    & $\Lambda$CDM    &--  & 0.012    & 12173    & 0.68	&  6.70 & -109.4  & 14.5 & 67.8 & 12.4\\
   S1    & $\Lambda$FDM    & 1.0  & 0.110   & 16841   & 0.51	&  4.65 & -78.5 & 8.5 & 198.3 & 7.9 \\
   S2    & $\Lambda$FDM    & 2.0  & 0.050   & 18235   & 0.45	&  4.88  & -79.4 & 9.0 & 147.3 & 8.3\\
   S3    & $\Lambda$FDM    & 4.0  & 0.031   &  17152  & 0.49	&  5.13 & -85.8  & 9.6 & 125.0 & 8.8\\
   S4    & $\Lambda$FDM    & 16.0  & 0.015   &  17253   & 0.50	&  5.70  & -93.0 & 11.0 & 90.0 & 9.9\\
\hline
\end{tabular}
\caption[]{The table shows the choice of parameters for the main scenarios considered in this work, as well as the main results. We fixed $f_{\rm esc}=0.1$ for all the main scenarios, but explore variations in Section \ref{sec:param}. The value of  $N_{\rm ion}$ corresponding to a model is estimated such that reionization ends at $z\sim6$. $N_{\rm ion}$ sets the fraction of population II stars ($f_{\rm PopII}$) considered in each model. We consider $S2$ be our fiducial $\Lambda$FDM model. We have used $f_X=1$ and $\alpha_X=0.5$ for these models. See the body of the text for details. \\
$^{\rm a}$ $\Delta^2_{\rm max}$ is evaluated at $k = 0.1 ~ h ~ \rm Mpc^{-1}$.}
\label{tab1}
\end{center}
\end{table}

\subsection{Main Scenarios}
\label{sec:mainscen}
In our four $\Lambda$FDM models we want to explore the impact of different values for $m_{22}$. We choose $m_{22}$=1, 2, 4 and 16. This parameter determines the halo mass function. We then calculate the value of $f_*$ which reproduces the star formation rate density between redshift 6 and 7 from \cite{MadauDickinson2014}. Finally we choose a value of $N_\mathrm{ion}$ which ensures that reionization finishes by $z=6$. The values of these parameters are listed in Table~\ref{tab1}. Note that the value of $N_\mathrm{ion}$ automatically implies a value for $f_\mathrm{PopII}$. The Thomson scattering optical depth for all models are consistent with the Planck range $\tau = 0.058 \pm 0.012$ from \cite{2016arXiv160503507P}.\footnote{The models are also consistent, to within $2\sigma$, with the tighter constraint on $\tau$ from Planck 2018: $\tau = 0.054 \pm 0.007$ \citep[$1\sigma$; ][]{Planck2018}. } We denote $S2$ to be our fiducial model for $\Lambda$FDM scenarios. For all the models we use $f_{\rm esc} = 0.1$, $f_X=1$ and $\alpha_X = 0.5$. 

We also include one $\Lambda$CDM scenario for which we used the same procedure to set the parameters choices. It uses a minimum halo mass for star formation $M_\mathrm{min}=M_{\mathrm{Ly}\alpha}$, see Eq.~\ref{Lyman alpha cooling}. This will make the $\Lambda$CDM results very different from the $\Lambda$FDM results just because of the difference in halo population. If the minimum mass of halos that can form stars would be much higher in the CDM case, the results could start to resemble the FDM ones. These kinds of degeneracies are hard to break, see for example the study of \cite{2014MNRAS.438.2664S} for the case of WDM. Our study does not attempt to identify unique features for FDM which would distinguish it from all possible CDM models but rather includes a CDM model for reference and shows what can be expected from FDM models.


\subsubsection{Global signal}
\label{global_si}
\begin{figure}
\begin{center}
\includegraphics[width=\textwidth]{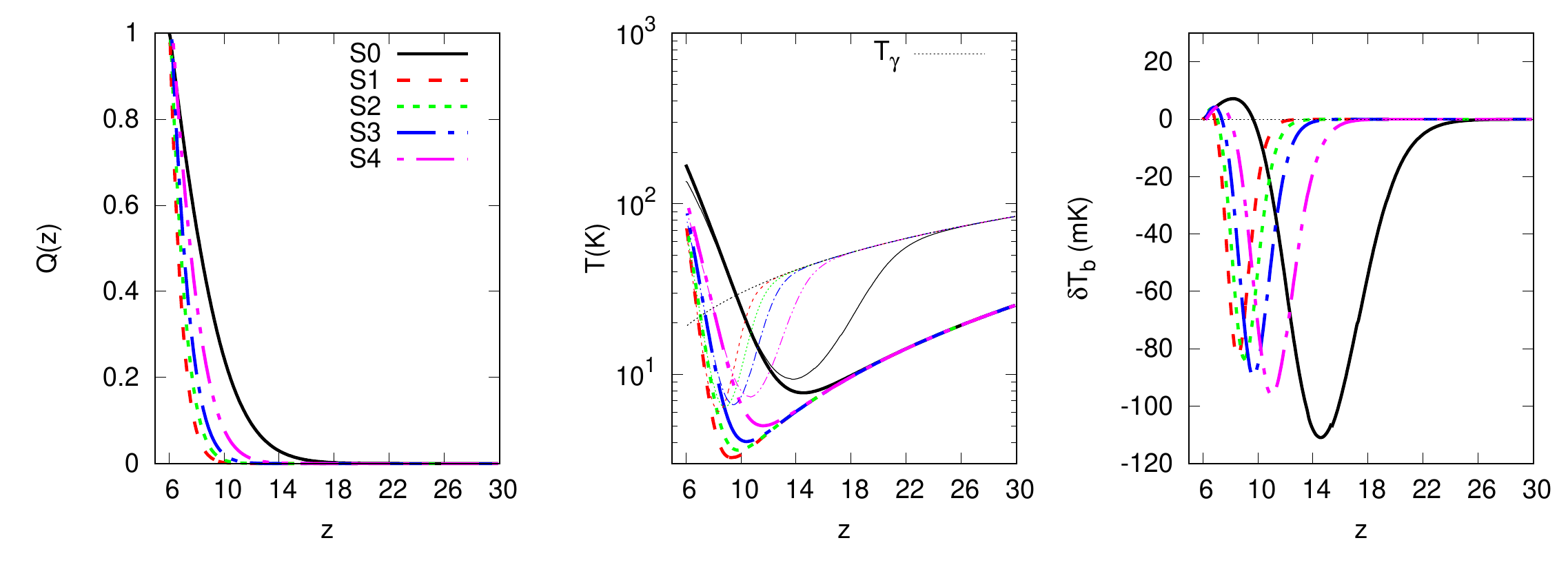}
    \caption{\textbf{Left-hand panel:} Redshift evolution of the volume averaged ionization fraction ($Q$) for different FDM models $S1$ (red dashed), $S2$ (green dotted), $S3$ (blue dash-dotted), $S4$ (magenta dash-double dotted ) and CDM model $S0$ (black solid). \textbf{Middle panel:} Redshift evolution of the average gas temperature  (thick) and spin temperature (thin) of the neutral IGM. The thin black dotted curve represent the redshift evolution of the CMB temperature. \textbf{Right-hand panel}: Differential brightness temperature of the 21-cm signal as a function of redshift for these models.}
   \label{image_p6globalxtk}
\end{center}
\end{figure}

First, we will discuss the impact of different  FDM models in terms of the global variables of EoR such as the volume averaged ionization fraction ($Q$), kinetic temperature of the partially ionized IGM and the mean brightness temperature. The redshift evolution of the volume averaged ionization fraction for these different models are shown in the left panel of Figure \ref{image_p6globalxtk}. 
We find the evolution of $Q$ is more rapid for the FDM model compared to the CDM model. For example, $Q$ evolves from 0.2 to 1 between the redshifts 8 and 6 for the FDM model $S2$, while this occurs between $z = 11$ and $z = 6$ for the CDM model $S0$. This is expected as the ionization history follows $f_{\rm coll}$ which evolves rapidly for the FDM models as we have seen previously. 

The evolution of the average $\TK$ for the different ${\Lambda}$FDM models as well as the CDM model are shown in the middle panel of Figure \ref{image_p6globalxtk}. One can easily notice that the heating process is severely delayed for the ${\Lambda}$FDM models compared to the ${\Lambda}$CDM model. For example, the average $\TK$ becomes larger than $\TCMB$ at $z \sim 11$ for the ${\Lambda}$CDM model, while this only happens at $z\sim 7.3$ for the $S2$ model.  The delay in the heating process is due to the delay in formation of the collapsed objects in the FDM models. As the number density of low mass haloes decrease with $m_{22}$ in the FDM models, the heating of the IGM occurs later for FDM models with smaller $m_{22}$ (e.g, compare models $S1$ and $S3$). In comparison to the CDM model, $\TK$ increases faster in the FDM models after $\TK$ reaches its minimum value. This is because $f_{\rm coll}$ increases faster in the FDM models compared to the CDM model.

Similar to the evolution of $Q$ and the average $\TK$ of the IGM, the $\lya$ coupling also follows the collapsed fraction. Thus, this process is also delayed for the FDM models compared to the CDM model. For example, $\TS$ follows $\TK$ from $z\sim 14$ for the CDM model, while same happens at $z\sim 9$ for the FDM model $S2$. In addition, the strength of the $\lya$ coupling is weaker for the FDM models and has a positive correlation with the value of $m_{22}$. 

All these effects are also reflected in the evolution of the average $\TB$ as shown in the right-hand panel of Figure \ref{image_p6globalxtk}. While the absorption signal for the CDM model becomes the strongest at $z\sim 16$, the 21-cm signal is negligible for all the FDM models at that redshift. On the other hand the CDM model shows an emission signal for $z\lesssim 12$ whereas the FDM models show a strong absorption signal in that regime. A lower value for $f_X$ in the CDM model would of course shift the absorption signal to lower redshifts. However, because the $\lya$ coupling is stronger in the CDM models, this absorption signal would be much stronger than what we see for the FDM cases.

Another way to delay the the X-ray heating process would be to lower the star formation efficiency $f_{\star}$ in the CDM models, which would also make the $\lya$ coupling weaker. However, in that case the absorption profile will be much wider compared to those from the FDM models. Thus, the growth rate of the average $\TB$ can in principle break the dark matter degeneracies \citep[also see][for the equivalent case for WDM]{2014MNRAS.438.2664S}.

The detection of a global 21-cm signal at redshifts $z\gtrsim 14$  will rule out or place strong constrains on the FDM models. Recently, \cite{2018Natur.555...67B} reported a detection of the absorption signal centered at a redshift of $z \simeq 17$ using low-band observations with EDGES. The depth of the absorption signal is much stronger than can be explained by standard physics and appears to require an unknown cooling agent operating at very high redshifts \citep{2018Natur.555...71B}. Combined with the spectral shape of the signal \citep{MirochaFurlanetto2018} as well as worries about the handling of the strong foreground signals \citep{Hills2018}, this raises considerable doubts about the reliability of the claimed result and confirmation by an independent group is required to give it strong credence. As it stands, however, this measurement disfavours our FDM models and would require FDM masses of $m_{22}\ge 50$ \citep{LidzHui2018} or even $m_{22}\ge 80$ \citep{2018PhRvD..98f3021S}. 

\cite{2017ApJ...847...64M} presented constraints from EDGES High-Band observations on the global 21-cm signal in the range $6<z<14$. Among these are constraints on the width of the absorption profile. For a depth between 50 and 100~mK, the EDGES observations disfavour Gaussian absorption profiles which have a FWHM of less than $\Delta z \sim 4$. FDM models S1--S4 in fact show absorption features which are narrower than this. This implies that these four models are ruled out by the EDGES High-Band observations.  This conclusion is of course dependent on the assumption that these constraints are reliable. In Section~\ref{sec:param} we will investigate a wider range of FDM models using this constraint.

\subsubsection{Bubble size distributions}
\label{sec:bub}
\begin{figure}
\begin{center}
\includegraphics[width=\textwidth]{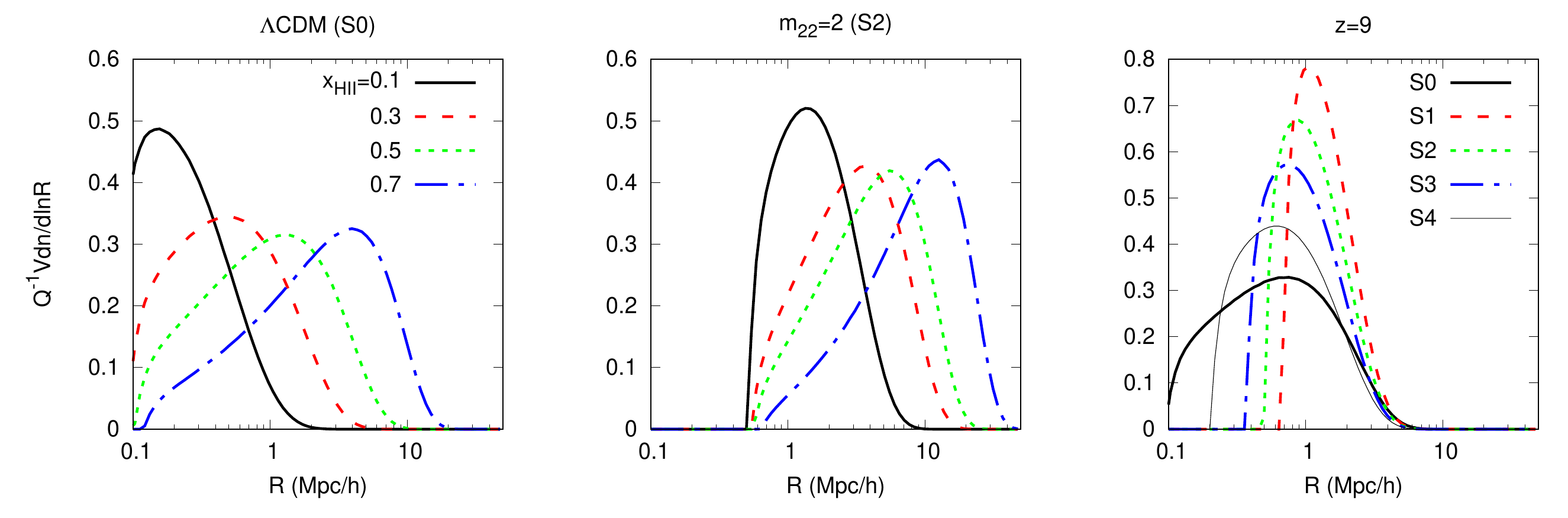}
   \caption{Bubble size distribution at different stages of reionization with ionization fraction 0.1 (solid), 0.3 (dashed), 0.5 (dotted) and 0.7 (dash-dotted). The left-hand and middle panels correspond to the CDM model $S0$ and FDM model $S2$ (fiducial model) respectively. The corresponding redshifts for $S0$ are 11.8, 9.5, 8.3 and 7.3 respectively, while these are 8.2, 7.3, 6.9 and 6.5 for $S2$. The right-hand panel represents the BSDs for all five models considered in this work at redshift 9. }
  \label{image_p6bub}
\end{center}
\end{figure}

The difference in source abundances between the FDM models and the CDM model lead to different bubble size distributions (BSDs). The BSDs for the CDM model $S0$ and FDM model $S2$ are shown in the left-hand and middle panels of Figure \ref{image_p6bub} respectively at different stages of reionization with ionization fraction 0.1, 0.3, 0.5 and 0.7. The most probable size of the ionised bubbles, as given by the peak of the BSDs, increases as reionization progresses. As one can easily notice, the most probable sizes of \HII ~regions are larger in the FDM models than in the CDM model when compared at the same stage of reionization, characterized here by the volume averaged ionization fraction. This is due to the fact that the FDM model lacks the low mass halos which produce small \HII ~regions. In fact the FDM BSD shows a clear smallest size which corresponds to the \HII ~region size produced by the lowest mass halo.

The right-hand panel of the figure present the BSDs for all 5 models considered in table \ref{tab1} at redshift 9. The BSDs from the FDM models are more peaked than the one from the CDM model and the lower $m_{22}$ the more peaked they become. In principle, if these BSDs can be measured in 21-cm tomographic imaging data \citep{giri2017bubble}, these differences will help break the degeneracies between the two dark matter models \citep{2017MNRAS.471.1936K, 2018arXiv180106550G}.


\subsubsection{Power spectra}
\label{ps_lb}
\begin{figure}
\begin{center}
\includegraphics[width=\textwidth]{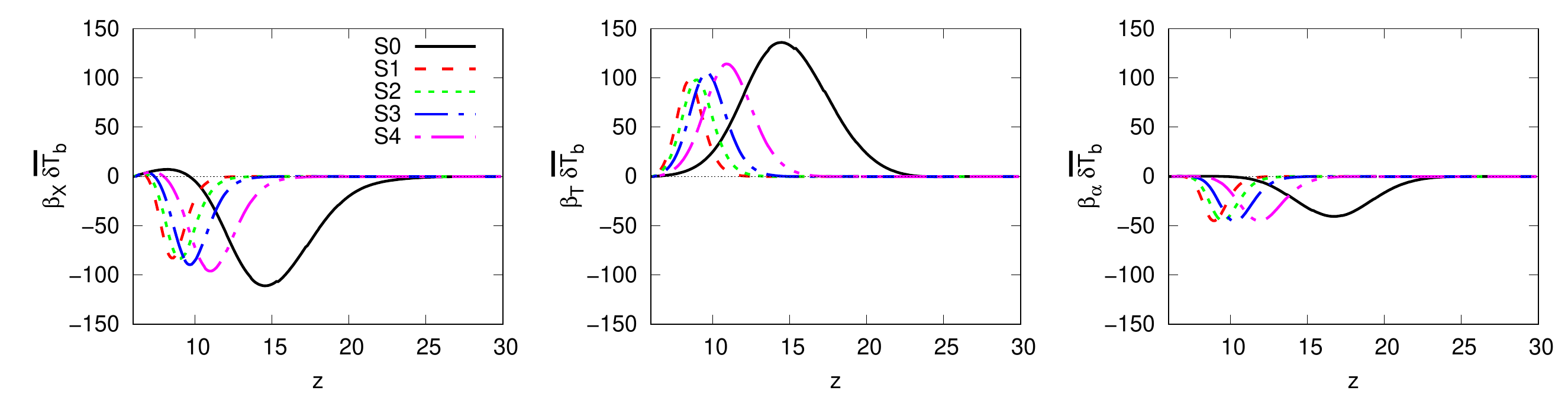}
    \caption{Redshift evolution of the fluctuations of the fundamental quantities for different FDM and CDM models. The left to right panels represent fluctuations due to ionization, gas temperature and $\lya$ coupling respectively. We do not plot the fluctuation from the baryonic density $\beta_{\rm B}\overline{\TB}$ as these follows the $\betaX \overline{\TB}$ curves.}
   \label{image_p6betas}
\end{center}
\end{figure}

Before discussing the power spectrum of the expected 21-cm signal from our five models, we will first study the individual fluctuation terms that contribute to the power spectrum (see equation \ref{eq:fluc}). As shown in Figure \ref{image_p6betas}, all models show that the 21-cm signal is dominated by the $\lya$ fluctuation initially, followed by $\TK$ and then $\XHI$ fluctuations. These fluctuations are weaker and delayed for the FDM models compared to the CDM models and also become weaker for smaller $m_{22}$ values.

\begin{figure}
\begin{center}
\includegraphics[width=\textwidth]{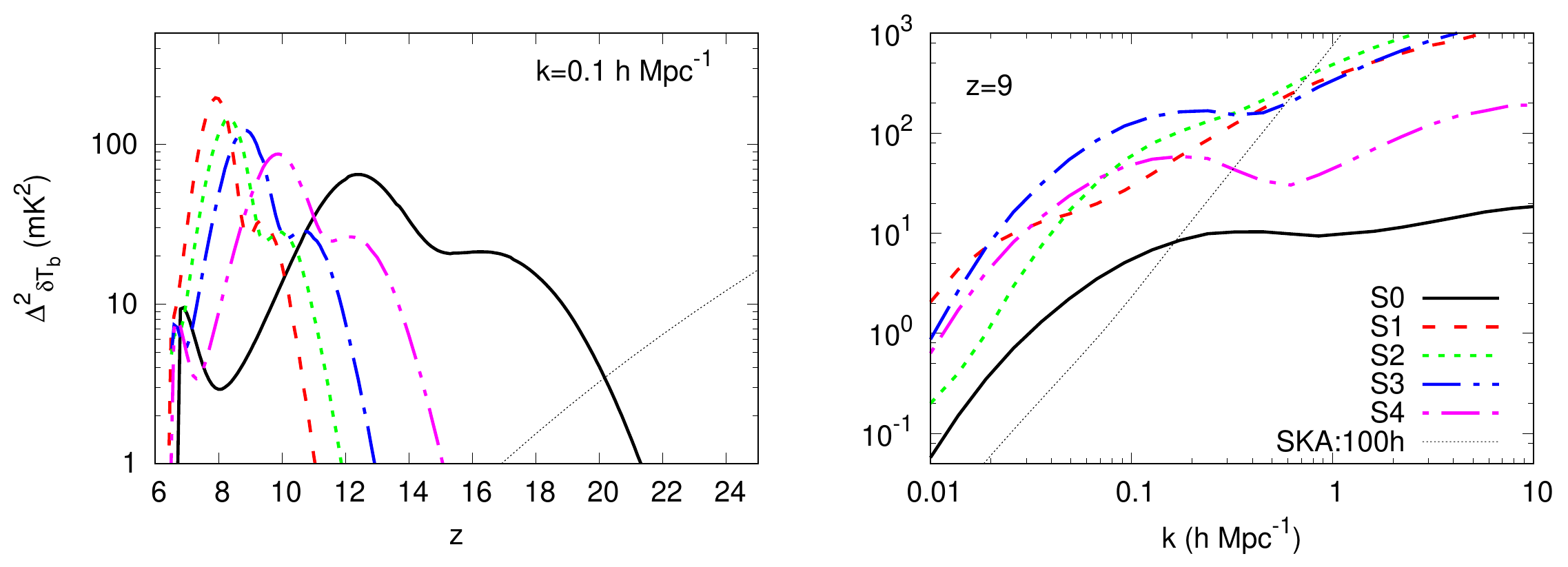}
    \caption{\textbf{Left-hand panel}: The redshift evolution of the power spectrum of $\TB$ at scale $k=0.1 ~h~\rm Mpc^{-1}$ for different models of FDM and CDM. The dotted curve corresponds to $1\sigma$ error on the power spectrum at scale $k=0.1 ~h~\rm Mpc^{-1}$  from the system noise from 100 h of observation with SKA1-low. We choose a bandwidth of 32 MHz and scale intervals $dk=k/5$ to estimate the thermal noise.  \textbf{Right-hand panel}: The power spectrum of $\TB$ at $z = 9$ as a function of scales for the selected FDM and CDM models.  The dips in the power spectrum at scales $\sim 0.5 ~h~\rm Mpc^{-1}$ for S$_3$ and S$_4$ occur as $\beta_T$ is negative whereas the other $\beta$ values are positive. Physically, this is due to the fact that the denser regions become hotter which eventually changes the signal from absorption into emission. See \cite{Pritchard07} for more details. The dotted curve corresponds to the same noise as shown in the left-hand panel at different scales.  } 
   \label{image_p6pstb}
\end{center}
\end{figure}

These effects are also visible in the evolution of the large scale ($k=0.1$~$h$ Mpc$^{-1}$) power spectrum of the 21-cm signal as shown in Figure \ref{image_p6pstb}. Note that the galaxy bias is higher for the FDM models which produce higher values of power spectrum compared to the CDM models. From the Cosmic Dawn to the end of reionization, different peaks of the curves in the left-hand panel of the figure correspond to $\lya$ coupling, heating and ionization fluctuations respectively. The effects of delayed $\lya$ coupling and X-ray heating in the FDM models are clearly visible by the shifts of the peaks. There is also a significant amount of overlap between the contributions from these fluctuations which makes the detection of these individual peaks more difficult for the FDM models. 

One can notice that the 21-cm signal is insignificant for the FDM models beyond redshift $\sim 14$, while the CDM model predicts a strong signal due to the $\lya$ and heating fluctuations. Thus, detection of a fluctuation signal at redshift $\gtrsim$14 will rule out the FDM models or provide strong constrains on the FDM models. The dotted thin curves correspond to 1-$\sigma$ error on the power spectrum due to thermal noise for 100 hours of observation time, 32 MHz bandwidth with SKA1-low and for intervals $\mathrm{d}k = k/5$. We follow \cite{2011MNRAS.418..516G, 2014JCAP...09..050V} for estimating the error due to thermal noise for 512 antenna of recent SKA1-low configuration.\footnote{The recent SKA1-low antenna configurations is taken from http://astronomers.skatelescope.org/} SKA1-Low observations at high redshifts should therefore be able to easily rule out or put strong constraints on FDM models.

Below $z\sim10$ the FDM models show a much stronger signal than the CDM model. The full power spectra at $z = 9$ are shown in the right-hand panel of Figure \ref{image_p6pstb} and reveal that this is true over a wide range of $k$ values. This implies that the detection of the 21-cm signal below $z = 10$  with telescopes such as LOFAR, MWA, HERA and SKA1-low will be easier if the background cosmology is driven by FDM. However, observations at multiple redshifts will be necessary to distinguish between FDM and CDM models with inefficient heating and or star formation. Also, as pointed out above, it will be challenging to distinguish CDM models in which star formation in low mass halos is suppressed from FDM models. In this sense it is easier to rule out than to positively confirm FDM models.


\subsection{Parameter space study}
\label{sec:param}
Above we showed the results for four distinct FDM scenarios in which we varied $m_{22}$ but kept $f_X$ and $\alpha_X$ constant. In this section we present a study in which we vary $f_X$, $\alpha_X$ and $m_{22}$ over a wide range so as to investigate the impact on the global signal as well as the power spectrum of the 21-cm signal. We choose  $f_X=1$, $\alpha_X=0.5$ and $m_{22}=2$ as our fiducial values. We vary two parameters at a time while the third parameter is fixed to its fiducial value. We characterize the results using the minimum value of $\TB$ as well as the maximum value of the large scale power spectrum, $\Delta^2 (k=0.1~h\rm Mpc^{-1})$, together with the redshifts when these  occur. These numbers provide a quantitative hint on the detectability of the signal in global experiments as well as with the interferometers.

\begin{figure}
\begin{center}
\includegraphics[width=\textwidth]{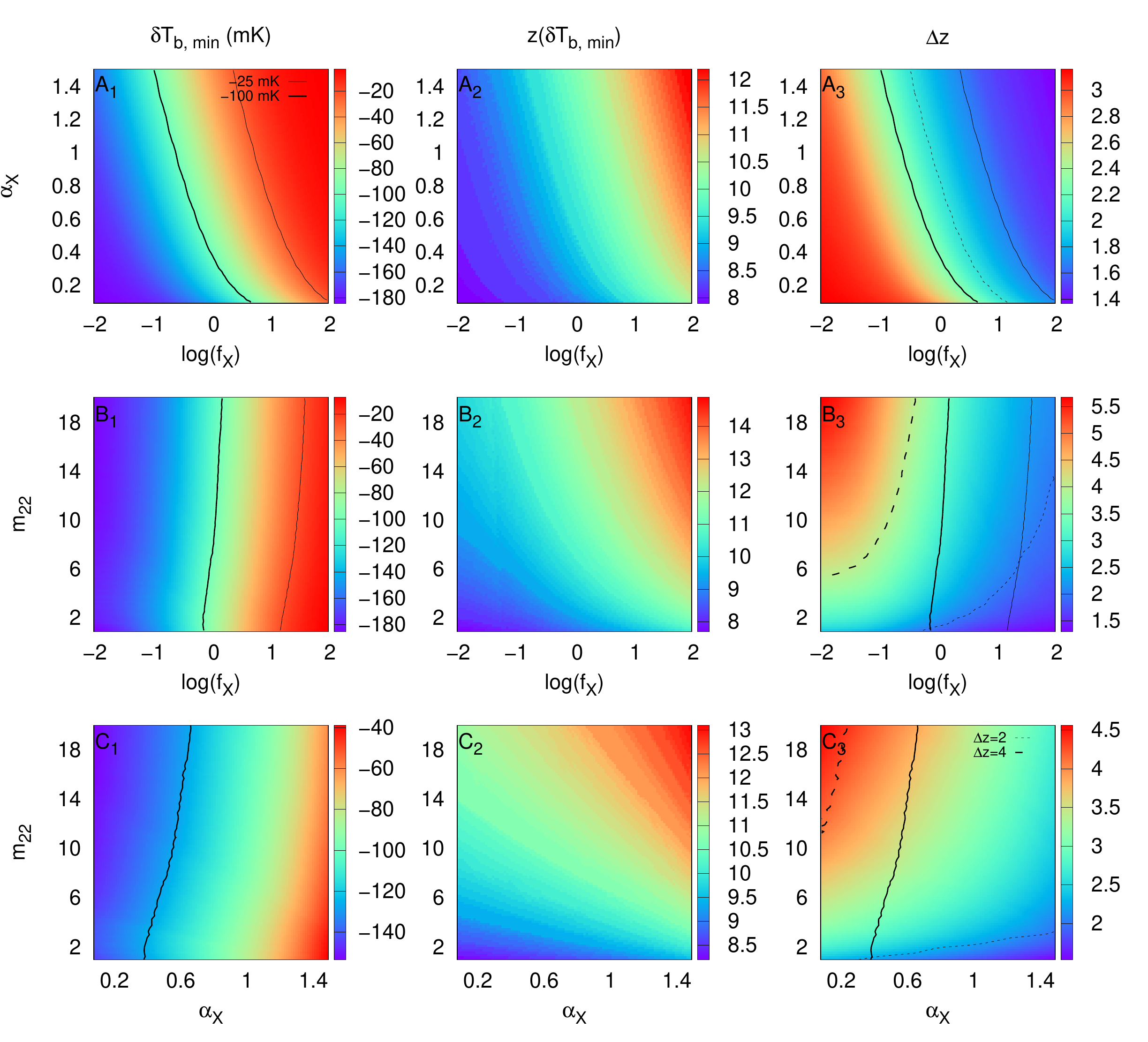}
   \caption{\textbf{Left-hand panels:} The minimum of the average brightness temperature (${\TB}_{, \rm min}$) throughout the reionization history. \textbf{Middle panels:} The corresponding redshift to ${\TB}_{, \rm min}$. \textbf{Right-hand panels:} The redshift width corresponding to the FWHM of the absorption profile of the brightness temperature.    We choose  $f_X=1$, $\alpha_X=0.5$ and $m_{22}=2$ as our fiducial parameter values. While we vary two parameters at a time, we fixed the third parameter to its fiducial value. }
  \label{fig:p6param}
\end{center}
\end{figure}

\begin{figure}
\begin{center}
\includegraphics[width=0.7\textwidth]{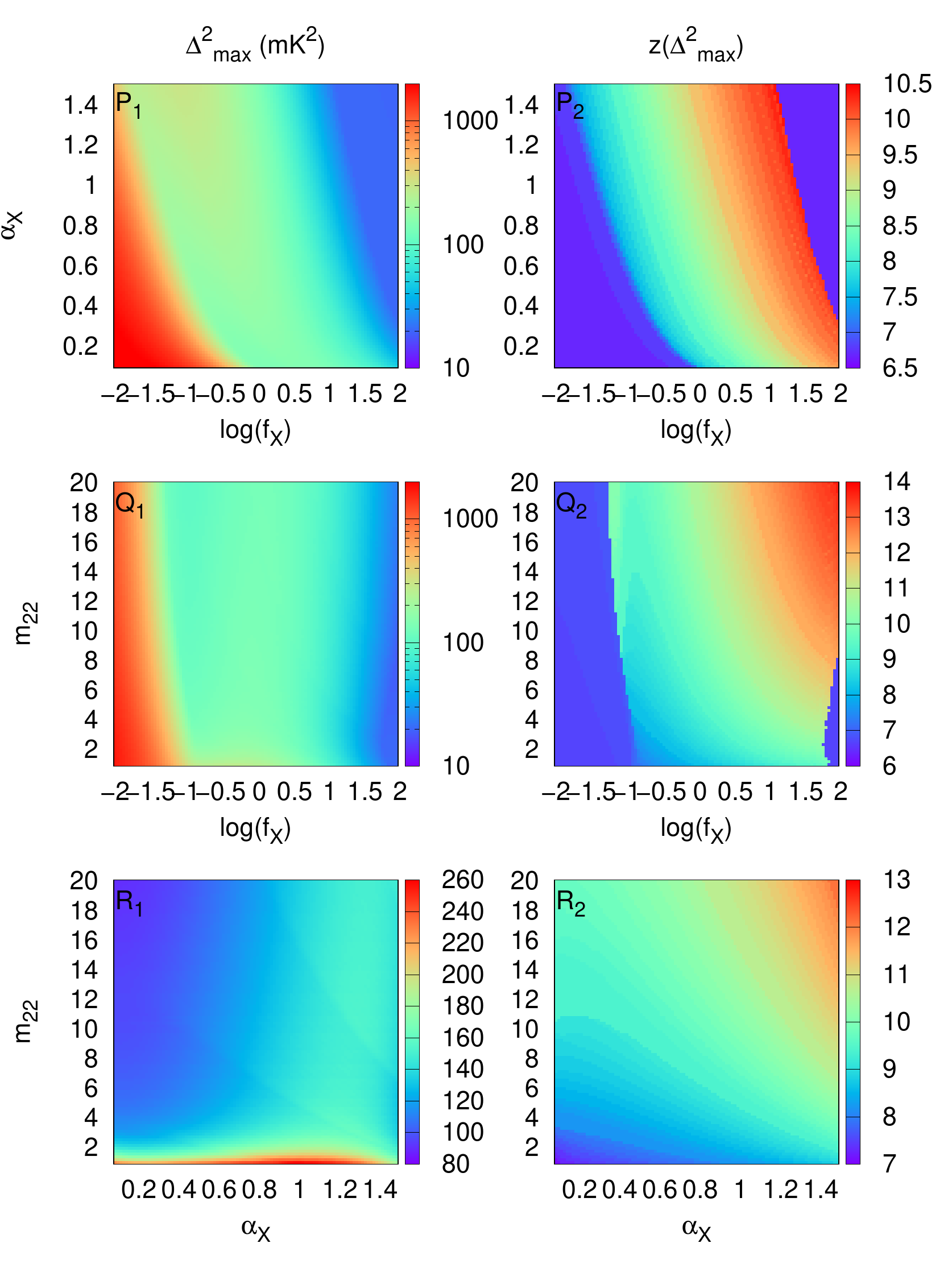}
   \caption{The left-hand panels present the maximum amplitude of power spectrum ($\Delta^2_{\rm max}$) at scale $k=0.1 ~ h ~ \rm Mpc^{-1}$ throughout the entire reionization period, while the right-hand panels present the corresponding redshifts to $\Delta_{\rm max}$. The fiducial choice of our parameters are  $f_X=1$, $\alpha_X=0.5$ and $m_{22}=2$. While we vary two parameters in each panel, we fixed the third parameter to its fiducial value. }
  \label{fig:p6paramps}
\end{center}
\end{figure}

The left-hand panels of Fig \ref{fig:p6param} present the minimum of the brightness temperature ${\TB}_\mathrm{, min}$, while the panels in the middle column show the associated redshifts $z({\TB}_\mathrm{, min})$. We see that ${\TB}_\mathrm{, min}$ is  more sensitive to the X-ray parameters $f_X$ and $\alpha_X$ than to $m_{22}$. The absorption signal during the Cosmic Dawn becomes weaker for larger values of $f_X$ as this leads to increased X-ray heating. Thus, the amplitude of the absorption trough (or the strongest absorption signal) decreases when $f_X$ increases as shown in the panel $A_1$ of the figure. The heating also occurs earlier for larger values of $f_X$, and thus, the position of the absorption trough shifts towards higher redshifts (see panel $A_2$ of the figure). The right-hand column of figure \ref{fig:p6param} represents the redshift width ($\Delta z$, full width at half maximum FWHM)  of the absorption profile around $z({\TB}_\mathrm{, min})$. The value of $\Delta z$ decreases with increasing $f_X$ and varies from 1.4 to 3.2 for $f_X\sim 100$ to $0.01$ respectively for the fiducial value of $m_{22}$. 

As the value of $\alpha_X$ in increased, the balance between soft and hard X-ray photons shifts more and more to soft X-rays. As soft X-rays more efficiently heat the neutral/partially ionised regions outside the \HII ~regions, heating become more efficient and occurs earlier when we increase $\alpha_X$. This results in a lower value for ${\TB}_\mathrm{, min}$ at earlier redshifts as can be seen in  panels $A_1$ and $A_2$. The same conclusion can also derived from panels $C_1$ and $C_2$ which show the ${\TB}_{\rm , min}$ and its corresponding redshift as a function of $m_{22}$  and $\alpha_X$ respectively. 

Panels $B_1$ and $B_2$ show that the value of ${\TB}_\mathrm{, min}$ depends weakly on the parameter $m_{22}$ within the explored range. However, ${\TB}_\mathrm{, min}$ decreases slightly when increasing $m_{22}$ as can be seen from panels $B_1$ and $C_1$. On the other hand, $z({\TB}_\mathrm{, min})$ shifts to higher redshifts when $m_{22}$ is increased.  This effect we had already seen in Fig.~\ref{image_p6globalxtk}. As the absorption signal becomes stronger for larger values of $m_{22}$, its FWHM increases. For example, for the fiducial values of $f_X$ and $\alpha_X$, $\Delta z$ changes from 2 to 3.8 when $m_{22}$ increases from 1 to 20.

As pointed out above \cite{2017ApJ...847...64M} constrained the FWHM for different values of ${\TB}_\mathrm{, min}$ using EDGES high-band data. Their study excluded models with $\Delta z$ smaller than 2 and 4 for ${\TB}_\mathrm{, min}$ deeper than -25 mK and -100 mK respectively using a Gaussian $\TB$ profile. The absorption profiles from our model can be well approximated by a Gaussian profile (see right-hand panel of Figure \ref{image_p6globalxtk}).The thin and thick solid curves in the left-hand and right-hand panels of figure \ref{fig:p6param} corresponds to the contours of -25 mK and -100 mK respectively, while the dotted and dashed curves in the right-hand panels of the figure represents $\Delta z$ equal to 2 and 4 respectively. For example, the parameter space between the curves corresponding to $\Delta z=4$ and ${\TB}_\mathrm{, min}$=-25 mK in panel $B_3$ is inconsistent with the results of \cite{2017ApJ...847...64M}. One can see that all the FDM models for the fiducial X-ray source in this study are in disagreement with their results.  

Based on the EDGES High-Band upper limits for $\Delta z$, our parameter study therefore implies that $m_{22}\gtrsim 6$ but then the X-ray efficiency has to be either very high or very low. For nominal X-ray efficiencies we find that $m_{22}> 20$ is needed. These constraints are more stringent than the ones previously derived on the basis of reionization histories \citep{Bozek2015,Corasaniti2017,Sarkar2016}, becoming comparable to constraints on $m_{22}$ from the $\lya$ forest for nomimal X-ray efficiencies  \citep{Armengaud2017, Irsic2017, LymanAlphaConstraints2017, Nori2019}. This shows the power of the 21-cm signal to constrain DM models.  As pointed out above, the results from the EDGES Low-Band observations have raised concerns about the reliability of the EDGES results. Even though the high-band results are of a different character than the low-band results, we should caution that the constraints on FDM obviously depend on the reliability of the claimed upper limits in \cite{2017ApJ...847...64M}. 

Interferometers such as LOFAR, MWA, HERA and the SKA will not be able to measure the global signal but are sensitive to the 21-cm power spectrum. We therefore also present results for this quantity. We chose to focus on the maximum value of the power spectrum $\Delta^2_{\rm max}$ at a scale $k=0.1~h ~ \rm Mpc^{-1}$ over the entire reionization history. All currently active and future interferometers are capable of measuring this scale. 
The left-hand panels of Fig.~\ref{fig:p6paramps} show the maximum value of $\Delta^2_{\rm max}(k=0.1~h ~ \rm Mpc^{-1})$ while the right-hand panels present the associated redshifts  $z(\Delta^2_{\rm max})$. 

Prediction of the maximum amplitude of the large-scale power spectrum and its position in redshift is not straightforward since the redshift evolution of the large-scale power spectrum shows three peaks, as described before. As astrophysical parameters are changed, the strongest peak may change from one of these three to another, leading to a sudden change of the redshift of the strongest peak. As one can see from panel $P_1$, similar to ${\TB}_\mathrm{, min}$, $\Delta^2_{\rm max}$ is also very sensitive to the X-ray parameters. For intermediate values of $f_X$, $\Delta^2_{\rm max}$ decreases slightly and appears towards higher redshift for higher values of $f_X$. However, for large values of $f_X\sim 100$, the IGM is heated very early leading to a smaller $\Delta^2$. In these cases, $\Delta^2_{\rm max}$ corresponds to the ionization peak and thus appears in the redshift range 6--7. Also when $f_X$ is very low ($\sim 0.01$), the ionization peak is stronger than the heating peak, see panel $P_1$ and $P_2$. 

Panels $R_1$ and $R_2$ show that for larger values of $\alpha_X$, $\Delta^2_{\rm max}$ increases slightly and shifts towards higher redshifts. This can be understood as an increase of the ratio of the number of soft X-rays to hard X-rays enhances the inhomogeneity of the heating process. On the other hand, $\Delta^2_{\rm max}$ decreases and appears later for higher values of $m_{22}$ which is consistent with Fig \ref{image_p6pstb}. 

LOFAR is capable of measuring the 21-cm power spectrum at scales of $k = 0.1 ~ h ~ \rm  Mpc^{-1}$ for redshifts below 11. The results from Fig.~\ref{fig:p6paramps} show that for a large part of the  parameter space explored here FDM models show appreciable power at redshifts accessible to LOFAR. Depending on the upper limits that LOFAR will be able to reach, these may lead to further constraints on the mass of the FDM particle.

\section{Summary and Conclusion}
\label{sec:conclude}
In this study, we have examined the evolution of the redshifted 21-cm signal from Cosmic Dawn and EoR in fuzzy dark matter cosmologies. For this we use an analytic model which incorporates the effects of $\lya$ coupling, X-ray heating and ionization to generate the ionization history and the expected 21-cm signal. As far as we know this is the first study for FDM incorporating all of these effects and considering the entire Cosmic Dawn and reionization epoch. Here we summarize our main findings.

\begin{itemize}
\item Compared to standard $\Lambda$CDM models, the severe reduction in the number of low-mass halos in $\Lambda$FDM models leads to a smaller number of the collapsed objects hosting ionizing sources. As a consequence the rate of ionizing photons per baryon produced by the sources must be high to ensure a reionization history which is both consistent with the CMB observations such as by Planck and a completion of reionization by $z=6$. We find that the required rates imply a significant contribution from PopIII stars assuming standard star formation efficiency and escape fraction values. The lower the value of $m_{22}$, the higher the required photon per baryon rate and the larger the contribution from PopIII stars.

\item Compared to standard $\Lambda$CDM models, there is also a considerable delay in the formation of collapsed objects which host ionizing sources. When requiring the end of reionization to be at $z=6$, this has as a result that the mean ionization fraction of the Universe evolves more rapidly in FDM models, as do the $\lya$ coupling and X-ray heating. The lower the value of $m_{22}$, the more rapid the evolution. Furthermore, whereas it is possible, although not necessary, to separate the three eras of $\lya$ coupling, X-ray heating and ionization in $\Lambda$CDM models, there will always be considerable overlap between them in FDM models. This implies that the assumption of spin temperature saturation in such models, as was for example used by \cite{Sarkar2016}, is not valid.

\item The redshift evolution of the globally averaged 21-cm signal is delayed in $\Lambda$FDM scenarios relative to $\Lambda$CDM. As a consequence the absorption profile is narrower and shallower. Narrower absorption profiles are easier to detect in global signal experiments. In fact, \cite{2017ApJ...847...64M} used results from the EDGES High-Band experiment to constrain the width of the absorption profile. For the parameter range which we explored these constraints translate into a lower limit for the mass of the FDM particle of $m_{22}\gtrsim 6$ but only for either very low or very high X-ray efficiencies. For nominal values of the X-ray efficiency $m_{22}$ would need to be higher than 20. 

\item Another consequence of the delay in the redshift evolution of the globally averaged 21-cm signal is that the 21-cm signal is very weak above a redshift $z \simeq 15$ for $m_{22}\leq 20$. Therefore a corroborated detection of the signal for $z>15$ from ongoing global signal detection experiments such as EDGES, SARAS, and LEDA will be able to put strong constraints on the dark matter particle mass $\mFDM$ in FDM scenarios. In fact, the claimed detection of an absorption signal at $z \simeq 17$ by the EDGES low band experiment \citep{EDGES2018} has already been shown to imply $m_{22}\geq 50$ \cite{LidzHui2018}.

\item The bubble size distribution for FDM cases differs considerably from the CDM case. The relative lack of small \HII ~bubbles in the FDM models lead to a narrower bubble size distribution. 

\item The evolution of the large scale 21-cm power spectrum in the FDM models resembles that of the CDM models, but just as for the global signal the features shift to lower redshifts and the evolution is more rapid. One consequence of the latter will be that for the FDM case the light cone effect \citep[see e.g.,][]{ghara15b} will have a stronger impact on the 21-cm signal power spectra than for the typical CDM case.

\item As the maximum of the power spectrum, which generally correspond to the heating peak, occurs at lower redshifts in the FDM models, the detectability of the signal is expected to be higher for both ongoing and future experiments such as LOFAR and SKA. The detection of a power spectrum signal at high redshifts ($z\gtrsim 15$) may rule out or put strong constrain on the FDM models. 
\end{itemize}

We want to reiterate that the FDM models may be degenerate with certain parameter choices for CDM models. Specifically, a higher minimal virial temperature for halos to produce ionizing photons can introduce features in the CDM models which will resemble those seen in the FDM models. Breaking these degeneracies is not easy and needs further investigation. However, since the conclusions are based on the HMF and the mapping of halo mass to photon production is complex, it will always be easier to rule out FDM models than to positively confirm them. 

The FDM mass constraint derived from the EDGES High-Band results \cite{2017ApJ...847...64M}, $\mFDM > 6$, has strong implications for $\Lambda$FDM as a viable model in solving small-scale challenges facing $\Lambda$CDM. Explaining observed constant-density cores in dwarf spheroidal galaxies in the context of $\Lambda$FDM favour (at $2\sigma$) $\mFDM = 1.18^{+0.28}_{-0.24}$ or $\mFDM = 1.79^{+0.35}_{-0.33}$, depending on the data set used \cite{Chen2017}. This is inconsistent with the 21-cm constraints derived in this paper, showing that $\Lambda$FDM cannot solve the cusp-core problem. Similarly, $\Lambda$FDM with $\mFDM > 6$ is ill-equipped to fully explain the dearth of dwarf galaxies. This is due to the fact that, as explained in Section \ref{sec:Structure_formation}, the halo mass function is only significantly suppressed on mass scales below $\MJeansEq \simeq 4.4 \times 10^8 ~ (\mFDM/6)^{-3/2} ~ \MSUN$. For $\mFDM > 6$, this is below $M_{\rm Ly\alpha} \simeq 3.0 \times 10^9 ~ (1 + z)^{-3/2} ~ \MSUN$ --- the $T_{\rm vir} \simeq 10^4 ~ \rm K$ lower limit for star formation set by efficient $\lya$ cooling and the reionization of the IGM --- for redshifts at least up to $z \simeq 2.6$. Thus, the observed scarcity of dwarf galaxies in halos with masses $< \rm few \times 10^{9} ~ \MSUN$ in the local Universe would have to be indicative of baryonic processes rather than exotic FDM effects. In summary,  if the EDGES High-Band constraints are reliable, they entail the failure of $\Lambda$FDM in addressing the astrophysical problems that motivated it in the first place. This shows the potential of 21-cm observations for dark matter studies, but in view of the important implications for the viability of $\Lambda$FDM also argues for independent confirmation of the results claimed by the EDGES team.


We caution that our models have made use of the Press-Schechter (ST) HMF, rather than the more accurate Sheth-Tormen (ST) HMF \cite{ShethTormen2002}. This is done mainly for consistency with our modelling of the \HII~ bubble distribution (see Section \ref{sec:bub}), the derivation of which implicitly assumes the Press-Schechter form to arrive at an expression for the \HII~ bubble formation barrier. As our model only uses the total collapsed fraction, we are only sensitive to differences between the two models in this quantity. The ST HMF has a weaker exponential cut-off for high-mass halos (i.e. halo masses $M$ for which $\deltacrit(M)/\sigma(M,z) \gtrsim 1$) than the PS HMF. This results in a slightly larger collapse fraction at higher redshifts which could ease our constraints on $\mFDM$ somewhat.\footnote{The PS and ST collapse fractions are nearly identical at $z = 6$, but the deviation grow larger at higher redshifts. In the redshift range $9 < z < 11$, we find that the PS collapse fraction with $\mFDM = 6$ is comparable to the ST collapse fraction with $\mFDM \simeq 4$.}  However, this would probably be offset by a more accurate semi-analytical modelling of the HMF for $\Lambda$FDM as done by \cite{Du2017}, who found that simply applying the ST HMF with $\deltacritEdS \rightarrow \deltacrit(M)$ leads to an underestimate (by a factor $\sim 2 - 3$ judging from their Figure 3) of the halo mass below which there is a sharp cut-off to the HMF, as well as an overestimate of the value of $\partial n/\partial \ln M$ at the peak of the HMF. Correcting for these effects would again lower the collapse fraction and lead to constraints on $\mFDM$ closer to what is found here using the PS HMF (if not more severe). Furthermore, we note that, while more accurate than the PS collapse fraction, the ST collapse fraction \textit{overestimate} the actual collapse fraction derived from N-body simulations \citep[see e.g. Figure 3 in][]{Reed2007}. 

As discussed in Section~\ref{sec:Structure_formation}, there are `extreme' versions of FDM which do not exhibit the same severe lack of low-mass halos as the `vanilla' FDM model studied in this paper. As a consequence, it is likely that these `extreme' versions of FDM would avoid the 21-cm constraints derived here. However, as we argued in Section~\ref{sec:Structure_formation}, these versions assume improbable initial conditions, making FDM far less attractive from the perspective of particle physics.

Although the exact value of the constraint from 21-cm observations may depend on the details of for example the star formation model, it is obvious that the severely delayed structure formation combined with an end of reionization by $z \simeq 6$ will always push the redshift width of the absorption signal to values of $\Delta z \lesssim 4$. In fact, we would expect interesting constraints also for WDM models as they in this respect resemble FDM models. Future 21-cm observations of either the global signal or the power spectrum can therefore be expected to lead to important constraints on the nature of dark matter.

\section*{Acknowledgements}
We would like to thank T. Roy Choudhury and Kanan K. Datta for useful discussions on this work, Jonathan Pritchard and Yue Bin for input regarding the analytical modelling, and Sunny Vagnozzi and Luca Visinelli for helpful comments on axion-like dark matter. We acknowledge the support from Swedish Research Council grant 2016-03581. We have also used resources provided by the Swedish National Infrastructure for Computing (SNIC) (proposal number SNIC 2018/3-40) at PDC, Royal Institute of Technology, Stockholm.

\bibliographystyle{JHEP}
\bibliography{reference2018}

\end{document}